\documentclass[apj,iop,twocolappendix,numberedappendix]{emulateapj}

\usepackage{color}
\usepackage{amsmath}
\usepackage{multirow}

\newcommand{\vect}[1]{{\mathbf{#1}}}
\newcommand{\madot}{\dot{M}_\mathrm{acc}}
\newcommand{\dt}{\Delta t}
\newcommand{\dx}{\Delta x}
\newcommand{\ms}{M_\mathrm{sink}}
\newcommand{\ma}{M_\mathrm{acc}}
\newcommand{\mo}{M_\mathrm{out}}
\newcommand{\rs}{\vect{R}_\mathrm{sink}}
\newcommand{\ra}{\vect{R}_\mathrm{acc}}

\newcommand{\vs}{\vect{V}_\mathrm{sink}}
\newcommand{\va}{\vect{V}_\mathrm{acc}}
\newcommand{\vo}{\vect{V}_\mathrm{out}}
\newcommand{\po}{\vect{P}_\mathrm{out}}
\newcommand{\ls}{\vect{L}_\mathrm{sink}}
\newcommand{\lacc}{\vect{L}_\mathrm{acc}}
\newcommand{\lo}{\vect{L}_\mathrm{out}}

\newcommand{\ssink}{\vect{S}_\mathrm{sink}}
\newcommand{\rhosink}{\rho_\mathrm{sink}}
\newcommand{\rsink}{r_\mathrm{sink}}
\newcommand{\rout}{r_\mathrm{out}}
\newcommand{\thetaout}{\theta_\mathrm{out}}
\newcommand{\fmass}{f_\mathrm{m}}
\newcommand{\fa}{f_\mathrm{a}}
\newcommand{\lref}{L_\mathrm{ref}}
\newcommand{\jres}{J_\mathrm{res}}

\newcommand{\alphavir}{\alpha_\mathrm{vir}}
\newcommand{\sfe}{\mathrm{SFE}}
\newcommand{\sfrff}{\mathrm{SFR}_\mathrm{ff}}
\newcommand{\sigs}{\sigma_s}
\newcommand{\scrit}{s_\mathrm{crit}}
\newcommand{\eps}{\epsilon}
\newcommand{\fudge}{\phi_\mathrm{fudge}}
\newcommand{\phit}{\phi_t}
\newcommand{\phix}{\phi_x}
\newcommand{\ycut}{y_\mathrm{cut}}
\newcommand{\tff}{t_\mathrm{ff}}
\newcommand{\mach}{\mathcal{M}}
\newcommand{\cs}{c_\mathrm{s}}
\newcommand{\msol}{\mbox{$M_{\sun}$}}
\newcommand{\rsol}{\mbox{$R_{\sun}$}}
\newcommand{\g}{\mathrm{g}}
\newcommand{\cm}{\mathrm{cm}}
\newcommand{\km}{\mathrm{km}}
\newcommand{\pc}{\mathrm{pc}}
\newcommand{\AU}{\mbox{AU}}
\newcommand{\s}{\mathrm{s}}
\newcommand{\yr}{\mathrm{yr}}
\newcommand{\Gauss}{\mathrm{G}}
\newcommand{\bfu}{\mathbf{v}}
\newcommand{\bfB}{\mathbf{B}}

\slugcomment{draft \today}

\shorttitle{Modeling jet and outflow feedback}
\shortauthors{Federrath et al.}

\begin{document}

\title{Modeling jet and outflow feedback during star cluster formation}

\author{Christoph~Federrath\altaffilmark{1,2}, Martin~Schr\"on\altaffilmark{3,4}, Robi~Banerjee\altaffilmark{5}, \& Ralf~S.~Klessen\altaffilmark{4}}
\email{christoph.federrath@monash.edu}

\altaffiltext{1}{Monash Centre for Astrophysics, School of Mathematical Sciences, Monash University, VIC~3800, Australia}
\altaffiltext{2}{The Australian National University, Canberra, ACT~2611, Australia}
\altaffiltext{3}{Department of Computational Hydrosystems, Helmholtz Centre for Environmental Research - UFZ, Permoserstr.~15, D-04318 Leipzig, Germany}
\altaffiltext{4}{Universit\"at Heidelberg, Zentrum f\"ur Astronomie, Institut f\"ur Theoretische Astrophysik, Albert-Ueberle-Strasse~2, D-69120 Heidelberg, Germany}
\altaffiltext{5}{Hamburger Sternwarte, Gojenbergsweg~112, D-21029 Hamburg, Germany}

% <~ 250 words
\begin{abstract}
Powerful jets and outflows are launched from the protostellar disks around newborn stars.
These outflows carry enough mass and momentum to transform the structure of their parent molecular cloud and to potentially control star formation itself.
Despite their importance, we have not been able to fully quantify the impact of jets and outflows during the formation of a star cluster. The main problem lies in limited computing power. We would have to resolve the magnetic jet-launching mechanism close to the protostar and at the same time follow the evolution of a parsec-size cloud for a million years. Current computer power and codes fall orders of magnitude short of achieving this.
In order to overcome this problem, we implement a subgrid-scale (SGS) model for launching jets and outflows, which demonstrably converges and reproduces the mass, linear and angular momentum transfer, and the speed of real jets, with $\sim1,\!000$ times lower resolution than would be required without SGS model.
We apply the new SGS model to turbulent, magnetized star cluster formation and show that jets and outflows (1) eject about 1/4 of their parent molecular clump in high-speed jets, quickly reaching distances of more than a parsec, (2) reduce the star formation rate by about a factor of two, and (3) lead to the formation of $\sim1.5$ times as many stars compared to the no-outflow case. Most importantly, we find that jets and outflows reduce the average star mass by a factor of $\sim3$ and may thus be essential for understanding the characteristic mass of the stellar initial mass function.
\end{abstract}


\section{Introduction}

Observations show that jets and outflows are launched from virtually all protostar--disk systems \citep[e.g.,][]{RicherEtAl2000,BeutherEtAl2002,WoitasEtAl2002,ShepherdTestiStarkAl2003,CabritEtAl2007,SwiftWelch2008,DuarteCabralEtAl2013,WangEtAl2014IRAS,DunhamEtAl2014}. And ALMA has now begun to push the resolution and detail of these observations to the next level \citep[e.g.,][]{SakaiEtAl2013,MurilloEtAl2013,MerelloEtAl2013}, providing us with important constraints for understanding and modeling the magnetic jet-launching process in computer simulations. Magnetic pressure-driven jets can even be reproduced in terrestrial experiments and the results of the experiments can be scaled to astrophysical dimensions \citep{CiardiEtAl2007,SuzukiVidalEtAl2010}.

Jets and outflows have been proposed to solve the following key problems in star formation research. First, they may help to explain the low star formation rate (SFR) and low star formation efficiency (SFE) in turbulent clouds \citep{PadoanEtAl2014,Krumholz2014,KrumholzEtAl2014}. Second, they contribute to solving the problem of efficiently removing angular momentum from the rotating, star-forming core and disk to explain the observed relatively slow rotation rates of stars \citep{PudritzEtAl2007,FrankEtAl2014}. Third, they could play a key role in explaining the observed stellar initial mass function (IMF) \citep{Chabrier2005,KroupaEtAl2013}, by removing some fraction of the accreting protostellar envelope, which may be parametrized with a core-to-star efficiency parameter $\eps\sim0.25$--$0.5$ \citep{MatznerMcKee2000,AlvesLombardiLada2007,NutterWardThompson2007,EnochEtAl2008,Myers2008,AndreEtAl2010,KoenyvesEtAl2010,FederrathKlessen2012,OffnerEtAl2014}.

Moreover, the kinetic energy injection rate of outflows and jets is of the order of the turbulent dissipation rate \citep{MacLowKlessen2004,KrumholzEtAl2014}, suggesting that outflows and jets can significantly contribute to driving and sustaining the observed level of turbulence \citep{SwiftWelch2008,MauryAndreLi2009,GravesEtAl2010,ArceEtAl2010,NakamuraEtAl2011,RivillaEtAl2013} and thus supporting a picture in which star formation might regulate itself by outflow feedback \citep[e.g.,][]{NormanSilk1980,Draine1983,Franco1984,ShuAdamsLizano1987,McKee1989}. Although numerical experiments by \citet{BanerjeeKlessenFendt2007} suggested that jets cannot generate supersonic turbulence if they are driven into a uniform-density medium at rest, jets and outflows may well be capable of sustaining and enhancing preexisting molecular cloud turbulence \citep{CunninghamEtAl2009,CarrollEtAl2009,CarrollFrankBlackman2010}. Even though outflows and jets are driven by magnetic pressure \citep{LyndenBell2003} and magneto-centrifugal acceleration \citep{BlandfordPayne1982} on sub-AU scales during protostar formation, their power can reach molecular clump and cloud scales, up to several parsecs \citep{TafallaMyers1997,ArceGoodman2001,StojimirovicEtAl2006,NarayananSnellBemis2012}. Considering the energy and scale of these outflows suggests that they cannot be ignored and likely have a strong impact on cloud dynamics and star formation.

Despite their potentially crucial impact on molecular cloud structure, dynamics and star formation, we have not been able to fully quantify the effects of jets and outflows. The main problem is to properly incorporate jets and outflows in a numerical simulation of star cluster formation, which is extremely challenging. We would have to resolve the magnetic launching mechanism, which occurs close to the protostar on sub-AU scales, while at the same time, we must follow the evolution of the star-forming cloud on parsec-size scales. A fully consistent calculation would require the numerical code to cover 7--8 orders of magnitude in length scales. Although calculations of \emph{single} stars have achieved such a high spatial resolution \citep[e.g.,][]{BanerjeePudritz2006,MachidaEtAl2008,BateTriccoPrice2014}, the time scale covered by those calculations is less than $2\,\mathrm{years}$ after protostar formation, even in the best state-of-the-art simulations. But forming a whole star cluster takes $10^5$--$10^6\,\mathrm{years}$ and requires us to follow \emph{multiple} sites of star formation at the same time. In a sentence: current codes and supercomputers are orders of magnitude away from solving this problem from first principles. A subgrid-scale (SGS) model is required that correctly reproduces the large-scale effects of jets and outflows, even if the actual launching mechanism is not resolved.

The aim of this study is \emph{i}) to implement an SGS model for driving jets and outflows during star cluster formation, which is based on theoretical, observational and numerical facts, and demonstrably reproduces the mass, linear momentum, angular momentum transfer, and speed of real jets and outflows, \emph{ii}) to determine the role of jets and outflows for the SFR, and \emph{iii}) to measure their impact on the characteristic mass of the IMF.

In Section~\ref{sec:methods}, we summarize the basic numerical techniques, including magnetohydrodynamics, gravity and sink particles. Section~\ref{sec:SGSmodel} presents the details of our new SGS outflow model. In Section~\ref{sec:tests}, we test the new model with isolated protostar formation calculations and demonstrate that it yields the converged mass, momentum, angular momentum transfer, and speed of real jets and outflows. Previous implementations of SGS outflow feedback and their limitations are discussed in Section~\ref{sec:discussion}. In Section~\ref{sec:cluster}, we apply our new SGS model to turbulent star cluster formation and determine the impact of outflow feedback on the SFR and on the characteristic mass of the IMF. Section~\ref{sec:conclusions} summarizes our conclusions.

\section{Basic numerical methods} \label{sec:methods}

Our SGS model for launching outflows is strictly bound to the sink particles used to model star formation and accretion. Here, we briefly describe the magnetohydrodynamics (MHD) scheme and the most important details of our sink particle algorithm, followed by a discussion of the coupling between the sink particles and our implementation of the SGS outflow model in Section~\ref{sec:SGSmodel} below.

\subsection{Magnetohydrodynamics and equation of state}

We use a modified version of the adaptive mesh refinement \citep[AMR,][]{BergerColella1989} code FLASH \citep{FryxellEtAl2000,DubeyEtAl2008} (in version 4) to integrate the three-dimensional, ideal MHD equations, including self-gravity,
\begin{eqnarray} \label{eq:mhd}
& & \frac{\partial \rho}{\partial t} + \nabla\cdot\left(\rho \bfu\right)=0\,, \nonumber\\
& & \rho\left(\frac{\partial}{\partial t} + \bfu\cdot\nabla\right)\bfu = \frac{(\bfB\cdot\nabla)\bfB}{4\pi} - \nabla P_\mathrm{tot} + \rho{\bf g}\,, \nonumber\\
& & \frac{\partial E}{\partial t} + \nabla\cdot\left[\left(E+P_\mathrm{tot}\right)\bfu - \frac{\left(\bfB\cdot\bfu\right)\bfB}{4\pi}\right] = \rho\bfu\cdot{\bf g}\,, \nonumber\\
& & \frac{\partial \bfB}{\partial t} = \nabla\times\left(\bfu\times\bfB\right)\,,\quad\nabla\cdot\bfB = 0\,,
\end{eqnarray}
where the gravitational acceleration of the gas ${\bf g}$, is the sum of the self-gravity of the gas and the contribution of sink particles (see Section~\ref{sec:sinks} below):
\begin{eqnarray} \label{eq:grav}
& & {\bf g} = -\nabla\Phi_\mathrm{gas} + {\bf g}_\mathrm{sinks}\,, \nonumber\\
& & \nabla^2\Phi_\mathrm{gas} = 4\pi G\rho\,.
\end{eqnarray}
Here, $\rho$, $\bfu$, $P_\mathrm{tot}=P_\mathrm{th}+ 1/(8\pi)\left|\bfB\right|^2$, $\bfB$, and $E=\rho \epsilon_\mathrm{int} + (\rho/2)\left|\bfu\right|^2 + 1/(8\pi)\left|\bfB\right|^2$ denote the gas density, velocity, pressure (thermal plus magnetic), magnetic field, and total energy density (internal plus kinetic, plus magnetic), respectively.

To model the thermal evolution during star formation, we use a polytropic equation of state
\begin{equation} \label{eq:eos1}
P_\mathrm{th}=K\rho^\Gamma,
\end{equation}
approximating the detailed radiation-hydrodynamic simulations by \citet{MasunagaInutsuka2000}. This covers the phases of isothermal contraction, adiabatic heating during the formation of the first and second core and the effects of $\mathrm{H}_2$ dissociation in the second collapse. According to these phases, we set the polytropic exponent in Equation~(\ref{eq:eos1}) to
\begin{equation} \label{eq:eos2}
\def\arraystretch{1.5}
\Gamma = \left\{
   \begin{array}{ll}
      1   & \,\mbox{for}\;\, \phantom{\;\rho_1 \leq} \rho \leq \rho_1 \equiv 2.50\times10^{-16}\,\g\,\cm^{-3}\,, \\
      1.1 & \,\mbox{for}\;\, \rho_1 < \rho \leq \rho_2 \equiv 3.84\times10^{-13}\,\g\,\cm^{-3}\,, \\
      1.4 & \,\mbox{for}\;\, \rho_2 < \rho \leq \rho_3 \equiv 3.84\times10^{-8\phantom{0}}\,\g\,\cm^{-3}\,, \\
      1.1 & \,\mbox{for}\;\, \rho_3 < \rho \leq \rho_4 \equiv 3.84\times10^{-3\phantom{0}}\,\g\,\cm^{-3}\,, \\
      5/3 & \,\mbox{for}\;\, \phantom{\;\rho_1 \leq} \rho > \rho_4\,.
   \end{array} \right.
\end{equation}
The polytropic constant $K$ in Equation~(\ref{eq:eos1}) is adjusted such that $K=\cs^2$, i.e, the square of the sound speed. When the polytropic exponent $\Gamma$ changes according to the density regimes given by Equation~(\ref{eq:eos2}), the temperature and sound speed change, and $K$ is adjusted, such that the pressure and temperature are continuous functions of density. In the isothermal regime ($\Gamma=1$), which serves as the normalization, the sound speed $\cs=0.2\,\km\,\s^{-1}$ and the temperature $T=11\,\mathrm{K}$ for gas with a typical molecular weight of $2.3\,m_\mathrm{H}$ (with $m_\mathrm{H}$ being the mass of a hydrogen atom). The initial isothermal evolution for $\rho\leq2.5\times10^{-16}\,\g\,\cm^{-3}$ is a reasonable approximation for dense, molecular gas of solar metallicity, over a wide range of densities \citep{WolfireEtAl1995,OmukaiEtAl2005,PavlovskiSmithMacLow2006,GloverMacLow2007a,GloverMacLow2007b,GloverFederrathMacLowKlessen2010,HillEtAl2011,HennemannEtAl2012,GloverClark2012}. We emphasize that Equations~(\ref{eq:eos1}) and~(\ref{eq:eos2}) are an approximation to hydrodynamic calculations that take radiative transfer effects into account \citep[e.g.,][]{Larson1969,YorkeBodenheimerLaughlin1993,MasunagaInutsuka2000,KrumholzKleinMcKee2007,OffnerEtAl2009,Bate2009rad,PetersEtAl2010,CommerconEtAl2010,MyersEtAl2013}. However, Equations~(\ref{eq:eos1}) and~(\ref{eq:eos2}) are a sufficient approximation for testing our SGS outflow model and even for cluster-formation simulations, as long as the opacity limit is resolved, which occurs at a density $\rho \gtrsim 10^{-14}\,\g\,\cm^{-3}$ \citep[e.g.,][and references therein]{Larson1969,Penston1969,Larson2005,JappsenEtAl2005}. If the opacity limit is resolved, the fragmentation of gas into stars with masses $M\gtrsim0.1\,\msol$ is roughly converged. Only the number of brown dwarfs is overestimated by factors of a few \citep[e.g.,][]{Bate2009cluster}.

All simulations use the positive-definite HLL3R Riemann scheme for ideal MHD \citep{BouchutKlingenbergWaagan2007,BouchutKlingenbergWaagan2010,Waagan2009,WaaganFederrathKlingenberg2011}, which has been tested for efficiency, robustness, and accuracy. \citet{WaaganFederrathKlingenberg2011} show that it maintains $\nabla\cdot\bfB\sim0$ with negligible errors. This MHD solver also allows us to model highly compressible gas flows including hypersonic jets and supersonic MHD turbulence without producing unphysical states. The self-gravity of the gas, i.e., the gas--gas gravitational interaction (Equation~\ref{eq:grav}) is computed with a multi-grid Poisson solver \citep[][]{Ricker2008} and the sink particle interactions are computed by direct $N$-body summation, as explained in Section~\ref{sec:sinks}. The gravitational potential and accelerations are computed according to the specified boundary conditions of the simulations.

We solve the MHD Equations~(\ref{eq:mhd}) and~(\ref{eq:grav}) in three dimensions with varying maximum effective resolutions $N_\mathrm{res}=2^{\lref}$ depending on the maximum AMR level $\lref$. On all AMR levels below the maximum, we refine the computational grid in regions where the Jeans length is resolved with less than 32 grid cells, in order to resolve turbulent vorticity and magnetic-field amplification on the Jeans scale \citep{SurEtAl2010,FederrathSurSchleicherBanerjeeKlessen2011,TurkEtAl2012}. Appendix~\ref{app:jeans} presents a Jeans-resolution study, demonstrating that we must resolve the Jeans length by more than 30 grid cells in order to achieve convergence. Scales smaller than the maximum refinement level are treated with sink particles and with our SGS model for launching jets and outflows.

\subsection{Sink particle formation and evolution} \label{sec:sinks}
In order to model collapse, accretion and star formation, we use sink particles \citep[for the first implementations of sink particles in smoothed particle hydrodynamics and in AMR, see][]{BateBonnellPrice1995,KrumholzMcKeeKlein2004}. In our implementation, only bound and collapsing gas forms stars and is allowed to be accreted \citep[for a detailed analysis and implementation, see][]{FederrathBanerjeeClarkKlessen2010}. The key feature of our approach is to define a control volume centered on grid cells exceeding a density threshold, $\rhosink$. \citet{TrueloveEtAl1997} found that the Jeans length must be resolved with at least 4 grid cells to avoid artificial fragmentation, leading to a resolution-dependent density criterion,
\begin{equation} \label{eq:rhosink}
\rhosink = \frac{\pi\cs^2}{4G\rsink^2}.
\end{equation}
The sink particle accretion radius $\rsink$ is typically set to 2.5 grid-cell lengths at the maximum level of refinement, sufficient to capture the formation and accretion accurately.

Grid cells exceeding the density threshold given by Equation~(\ref{eq:rhosink}), however, do not form sink particles right away. First, a spherical control volume with radius $\rsink$ is defined around the cell exceeding $\rhosink$, in which a series of checks for gravitational instability and collapse are performed \citep{FederrathBanerjeeClarkKlessen2010}. A sink particle is only created, if the gas in the control volume
\begin{enumerate}
\setlength{\itemsep}{1.0pt}
\item is on the highest level of grid refinement,
\item is not within $\rsink$ of an existing sink particle,
\item is converging from all directions ($v_r < 0$),
\item has a central gravitational potential minimum,
\item is bound ($|E_\mathrm{grav}|>E_\mathrm{thermal}+E_\mathrm{kin}+E_\mathrm{mag}$),
\item and is Jeans-unstable.
\end{enumerate}
This procedure avoids spurious sink particle formation, and allows us to trace only truly collapsing and star-forming gas.

Once a sink particle is created, it can accrete gas from the AMR grid, but only if the gas exceeds the density threshold, is inside the sink particle accretion radius, is bound to the particle, and is collapsing toward it. If all these criteria are fulfilled, the excess mass above the density threshold defined by Equation~(\ref{eq:rhosink}) is removed from the MHD system and added to the sink particle, such that mass, momentum and angular momentum are conserved by construction (as shown in detail in the next section).

The gravitational interaction between sink particles and with the gas is computed by direct $N$-body summation over all sink particles and grid cells. We use a second-order Leapfrog integrator to advance the sink particles with a velocity-based and acceleration-based timestep constraint that allows us to resolve close and highly eccentric orbits of sink particles without introducing any errors on super-resolution grid scales. Detailed tests of this method are provided in \citet{FederrathBanerjeeClarkKlessen2010} and \citet{FederrathBanerjeeSeifriedClarkKlessen2011}.

\subsection{Sink particle accretion}

The mass fraction $\Delta m_i$ to be accreted from cell $i$ with mass $m_i$ and volume $V_i$ is
\begin{equation}
\Delta m_i = m_i - \rhosink V_i\,.
\end{equation}

Within the control volume $(4\pi/3)\rsink^3$ of a sink particle, we gather the mass, center of mass (c.o.m.), momentum and angular momentum of the material to be accreted:
\begin{equation}
\def\arraystretch{1.5}
\setlength{\tabcolsep}{1.0pt}
\begin{tabular}{rrl}
\text{ mass: } & $\ma$ & $= \sum_i \Delta m_i$ \\
\text{ c.o.m.: } & $\ma \ra$ & $= \sum_i \Delta m_i \vect{r}_i$ \\
\text{ momentum: } & $\ma \va$ & $= \sum_i \Delta m_i \vect{v}_i$ \\
\text{ ang.~mom.: } & $\lacc$ & $= \sum_i \Delta m_i \vect{r}_i \times \vect{v}_i$\,.
\end{tabular}
\end{equation}

We then remove the mass $\Delta m_i$ from each cell and add the accreted material to the sink particle, in order to fulfill the conservation laws of the mass, center of mass, momentum, and total angular momentum (which is the sum of the sink particle's angular momentum $\ls$ and spin $\ssink$), denoting quantities after the accretion step with a prime:
\begin{equation}
\def\arraystretch{1.5}
\setlength{\tabcolsep}{1.0pt}
\begin{tabular}{rrl} \label{eq:sinkaccretion}
\text{ mass: } & $\ms'$ & $= \ms+\ma$ \\
\text{ c.o.m.: } & $\ms' \rs'$ & $= \ms \rs + \ma \ra$ \\
\text{ momentum: } & $\ms' \vs'$ & $= \ms \vs + \ma \va$ \\
\text{ ang.~mom.: } & $\ls'$ & $= \ms' \rs' \times \vs'$ \\
\text{ spin: } & $\ssink'$ & $= \ssink + \ls - \ls' + \lacc$\,.
\end{tabular}
\end{equation}

The sink particle spin $\ssink$ was introduced in order to absorb the excess accreted angular momentum. Note that this is not simply $\lacc$ as one might first think, because the sink particle position $\rs$ and thus the angular momentum of the sink particle $\ls$ can change slightly during an accretion event (in order to conserve the c.o.m.), which is compensated by the additional term $\ls - \ls'$, such that global angular momentum is conserved \citep[see Appendix~B in][]{FederrathBanerjeeClarkKlessen2010}.

Finally, we note that we do not directly modify the magnetic field when a sink particle is created or when it accretes. We carefully thought about accreting magnetic flux in addition to accreting mass, but we concluded that this would introduce two problems that can be avoided by leaving the magnetic field intact. The first problem arises when we would modify the magnetic field on the grid. This must be done such that the magnetic field remains divergence-free, i.e., $\nabla\cdot\vect{B}=0$. Although this may be achievable, the pressure exerted by the magnetic field would be lost when it is removed from the grid. This creates another problem, i.e., we would have to account for this loss of magnetic pressure, e.g., by adding artificial magnetic correction forces inside the sink particle (note that we also have to account for the gravity from the sink particle because of mass accretion, and the same would apply for effects of the magnetic field in the vicinity of the sink particles, if we accreted magnetic flux). Thus, instead of tempering with the magnetic field, we leave it intact and only accrete mass, such that the aforementioned problems are avoided. We note that numerical diffusion of the magnetic field still takes place on the grid scale, but this is a relatively small effect \citep[for a quantification of numerical diffusion of the magnetic field, see][Appendix]{FederrathSurSchleicherBanerjeeKlessen2011}, and happens independently of whether sink particles are included or not.

\section{The outflow/jet model} \label{sec:SGSmodel}

Due to the modular concept of the code, the outflow module applies when accretion has finished. This sequence is repeated in each time step of the code. This has the advantage that the state of the system after the accretion step is completely determined, and a well-defined fraction of the accreted material can be re-inserted to launch the outflow in the feedback step. This requires two loops over all sink particles and grid cells within the accretion and outflow radii. The accretion and outflow modules are structurally separated in the code, with the outflow module depending on the accretion module, but not vice versa.

\subsection{Geometry of the outflow launching region}

\begin{figure}
\centerline{\includegraphics[width=0.99\linewidth]{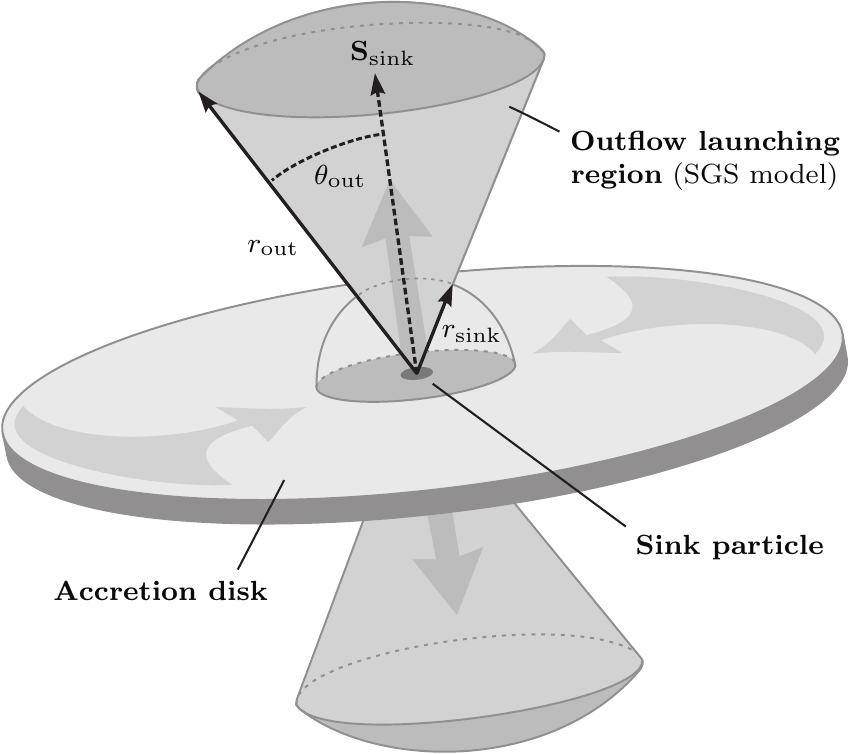}}
\caption{Schematic of our subgrid-scale outflow model, showing the basic geometry, with the sink particle radius $\rsink$ and spin $\ssink$, the outflow radius $\rout$ and the opening angle $\thetaout$. The arrows indicate gas flows in the disk and SGS region. Not to scale.}
\label{fig:schematic}
\end{figure}

We use two spherical sectors with radius $\rout$ around the sink particle in order to insert the outflow. Figure~\ref{fig:schematic} shows a schematic of our SGS outflow model, indicating the sink particle spin $\ssink$, the sink particle radius $\rsink$ and the outflow radius $\rout$. The outflow launching region is defined as the two spherical cones with the opening angle $\thetaout$ centered on the spin axis of the sink particle. In the following, we will refer to the outflow-launching region as the top and bottom outflow `cones'.

Observations and theoretical models show that outflows are typically aligned with the rotation axis of the accretion disk \citep[][]{AppenzellerMundt1989}, because the magnetic field is wound-up in the disk, creating a magnetic tower \citep{LyndenBell2003} along the rotation axis. Moreover, opening angles of $\thetaout=30^\circ$ relative to the rotation axis are consistent with magneto-centrifugal acceleration of the jet component \citep{BlandfordPayne1982}. As $\thetaout=30^\circ$ is the standard value for magnetically-driven disk winds and consistent with observations of protostellar outflows \citep[e.g.,][]{AppenzellerMundt1989,Camenzind1990}, we apply this opening angle as the default value in our SGS model.

\subsection{Mass transfer} \label{sec:masstransfer}

The key adaptive element of our SGS model is to reinsert a well-defined mass fraction of accreted material and launch that mass away from the sink particle to model an outflow+jet component. The outflow mass $\mo$ inserted in each time step $\dt$ is determined by the accretion rate of the sink particle $\madot$, according to
\begin{equation} \label{eq:mo}
\mo = \fmass \madot \dt.
\end{equation}
The mass fraction $\fmass$ can be chosen arbitrarily by the user of the SGS model, but theories based on the centrifugal acceleration mechanism and the X-wind model \citep[e.g.,][]{BlandfordPayne1982,PudritzNorman1986,ShuEtAl1988,WardleKoenigl1993,KoeniglPudritz2000,PudritzEtAl2007}, as well as observations \citep[e.g.,][]{HartmannCalvet1995,Calvet1998,BacciottiEtAl2002,CabritEtAl2007,LeeEtAl2006,LeeEtAl2007,BacciottiEtAl2011} suggest $\fmass\sim0.1$--$0.4$. For instance, \citet{BanerjeePudritz2006} find $\dot{M}_\mathrm{jet}/\madot\sim0.33$ for the inner jet component in their AMR simulations, which is also consistent with this range. Due to dust  obscuration, observations are restricted to measuring the accretion and outflow rates at relatively large distances from the accretion disk. Thus, the fraction $\fmass$ may be somewhat different closer to the source of the jet, inside the disk, but likely similar. \citet{SeifriedEtAl2012} find $\fmass\sim0.1$--$0.6$ in high-resolution simulations, because of spatial and temporal variations. Averaging over all space and time, \citet{SeifriedEtAl2012} find a mean value of $\fmass\sim0.3$, in agreement with other numerical simulations \citep{Tomisaka1998,Tomisaka2002,CasseKeppens2002,HennebelleFromang2008,DuffinPudritz2009,SheikhnezamiEtAl2012,FendtSheikhnezami2013}. The mass fraction $\fmass$ is remarkably insensitive to changes in the physical conditions of the disk and the progenitor core. The magnetic field strength and level of rotation do not seem to significantly change $\fmass$. Using smoothed particle magnetohydrodynamics simulations, \citet{PriceTriccoBate2012} recently arrived at a similar conclusion and found values of $\fmass$ up to $0.4$.

Based on the range of values for the mass fraction $\fmass$ inferred from theoretical models, observations, and numerical simulations, and based on the relatively weak dependence of $\fmass$ on the magnetic field and initial angular momentum of the star-forming core, we adopt $\fmass=0.3$ as the reasonable standard value. In Section~\ref{sec:tests}, we also study the dependence of our results on $\fmass$ and find that it does not significantly alter the resulting mass and momentum injection, because our SGS model is self-regulating.

In each timestep, we add the outflow mass $\mo$ given by Equation~(\ref{eq:mo}) uniformly to the gas within the outflow cones and subtract it from the sink particle, in consideration of mass conservation. In order to achieve a smooth transition at the interface of the SGS launching region, we use a radial and an angular smoothing function, defined as
\begin{eqnarray}
\mathcal{R}(r,\rout) & = &
\left\{
\def\arraystretch{1.4}
\begin{array}{cr}
\sin\left[\pi(r/\rout)\right] & \;\; \textnormal{for $r \leq \rout$}\\
0 & \textnormal{for $r > \rout$}
\end{array}
\right., \label{eq:radsmooth} \\
\Theta(\theta,\thetaout) & = &
\left\{
\def\arraystretch{1.4}
\begin{array}{cr}
\cos^p\left[(\pi/2)(\theta/\thetaout)\right] & \;\; \textnormal{for $\left|\theta\right|\leq \thetaout$}\\
0 & \textnormal{for $\left|\theta\right| > \thetaout$}
\end{array}
\right., \label{eq:angsmooth}
\end{eqnarray}
such that the effect of the SGS model quickly approaches zero toward the interface of the outflow cones. We have experimented with different choices of the smoothing functions and with different values of the smoothing power $p$ and did not find any significant effect on the shape of the outflow as long as $p\leq4$. For simplicity, we use $p=1$ as the standard value. The model works even without any smoothing toward the boundaries of the outflow cones, but in some rare cases, the sharp transition without smoothing can cause numerical instabilities, which we avoid by using the simple smoothing functions given by Equations~(\ref{eq:radsmooth}) and~(\ref{eq:angsmooth}).

\subsection{Momentum transfer} \label{sec:momtransfer}

Given the mass $\mo$ inserted in each timestep (Equation~\ref{eq:mo}), the momentum transferred to each of the two outflow cones in the rest frame of the sink particle is simply
\begin{equation} \label{eq:po}
\po = \pm \frac{1}{2} \mo \vo .
\end{equation}
For the radial outflow velocity $\vo$, we use the Kepler speed at the footpoint of a centrifugally-driven jet, close to the protostellar radius, as suggested by analytic models \citep{BlandfordPayne1982,ShibataUchida1985,ShibataUchida1986,PudritzNorman1986,WardleKoenigl1993,KoeniglPudritz2000}. The Kepler speed for a typical protostar with a mass of $M=0.5\,\msol$ at a radius of $R=10\,\rsol$ is $V_\mathrm{Kepler}=(GM/R)^{1/2}\sim100\,\km\,\s^{-1}$. This is indeed the typical outflow speed measured in observations \citep[e.g.,][]{Herbig1962,SnellLorenPlambeck1980,BacciottiEtAl2002}, so we use it to normalize our outflow model. The actual outflow speed, however, depends on the mass of the sink particle. For the outflow velocity we thus use
\begin{equation} \label{eq:vrad}
|\vo| = \left(\frac{G\ms}{10\,\rsol}\right)^{1/2} = 100\,\km\,\s^{-1}\;\left(\frac{\ms}{0.5\,\msol}\right)^{1/2}.
\end{equation}
Since the radius of a protostar can vary between $1\,\rsol$ and $100\,\rsol$ for stars within the wide mass range of \mbox{$0.1$--$100\,\msol$} and for a typical range of accretion rates of \mbox{$10^{-6}$--$10^{-3}\,\msol\,\yr^{-1}$} \citep{HosokawaOmukai2009}, we choose to normalize our SGS model with the Kepler speed of a protostar with an intermediate launching radius of $10\,\rsol$, about twice the radius of a solar-type protostar \citep{StahlerShuTaam1980a,StahlerShuTaam1980b,Larson2003}. Equation~(\ref{eq:vrad}) takes into account that the outflow accelerates with increasing mass of the protostar, which is indeed observed in dedicated outflow simulations without the SGS model, discussed below. The dependence of $\vo$ on the sink particle mass means that the outflow and jet turn on smoothly, similar to a real jet, gradually drilling through the accretion flow along the rotation axis of the disk.

\begin{figure}
\centerline{\includegraphics[width=0.9\linewidth]{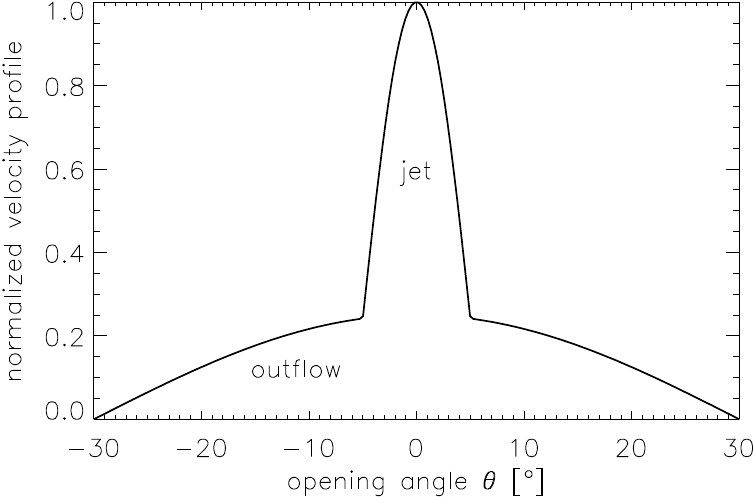}}
\caption{Normalized velocity profile of our SGS model, Equation~(\ref{eq:vprofile}), which consists of a low-speed, wide-angle outflow and a high-speed, collimated jet, as seen in previous simulations and in observations.}
\label{fig:velshape}
\end{figure}

Observations, theoretical models and simulations indicate that outflows typically consist of two components, a low-speed, wide-angle outflow and a high-speed, collimated jet \citep[][]{SnellLorenPlambeck1980,Draine1983,UchidaShibata1985,BacciottiEtAl2000,BanerjeePudritz2006,ShangEtAl2006,MachidaEtAl2008,AgraAmboageEtAl2011,VelusamyLangerThompson2014}. The outflow component might be driven from the first core by a magnetic tower flow, while the jet component is likely launched from the second core by magneto-centrifugal acceleration. In order to capture both components in an approximate way with our SGS model, we construct a normalized velocity profile $\mathcal{V}(\theta,\thetaout)$, consisting of a slow, wide-angle outflow and a fast, collimated jet \citep[see e.g.,][]{Camenzind1990,MachidaEtAl2008},
\begin{equation} \label{eq:vprofile}
\mathcal{V}(\theta,\thetaout)=\frac{1}{4}\Theta(\theta,\thetaout)+\frac{3}{4}\Theta(\theta,\thetaout/6),
\end{equation}
where $\Theta(\theta,\thetaout)$ is the angular smoothing function defined in Equation~(\ref{eq:angsmooth}) with $p=1$. For our fiducial opening angle $\thetaout=30^\circ$, we obtain a wide-angle ($30^\circ$), low-speed component with $\sim0.25\,\vo$ and a collimated ($5^\circ$), high-speed jet with $\vo$ \citep[for early observational studies of jets with collimation angles of $5$--$10^\circ$, see e.g.,][]{MundtFried1983,AppenzellerMundt1989}. Even terrestrial experiments of magnetic pressure-driven jets show such a high degree of collimation with angles $<10^\circ$ \citep{CiardiEtAl2007}. Figure~\ref{fig:velshape} shows the normalized velocity profile given by Equation~(\ref{eq:vprofile}), clearly depicting the two separate components. We have chosen the relative power and opening angles of the outflow and jet components to best resemble numerical simulations that can distinguish the two components \citep{BanerjeePudritz2006,MachidaEtAl2008}. The normalized velocity profile is multiplied with the radial outflow velocity $\vo$ from Equation~(\ref{eq:vrad}).

Finally, we note that we have paid particular attention to implementing the momentum transfer such that it is symmetric and conserves global momentum. It may seem trivial to achieve symmetry and momentum conservation, because the same amount of mass and momentum should be inserted in the top and bottom cones. Due to the discretization of the grid, however, and the fact that the position of the sink particle and the outflow axis can be arbitrarily oriented with respect to the AMR grid, this is not trivial at all and requires multiple loops over all grid cells within the outflow cones. First, we define the cells in the top and bottom cones and distribute the mass $\mo$ and momentum $\po$ in each cone. In a second loop, we correct for the slight deviation of mass and momentum, such that the top and bottom cones carry the same mass and transfer exactly opposite momenta. Doing this, we ensure that the total mass $\mo$ and the total momentum transferred to the outflow matches our target values given by Equations~(\ref{eq:mo}) and~(\ref{eq:po}).

\subsection{Angular momentum transfer} \label{sec:angmomtransfer}

Outflows and jets rotate \citep{BacciottiEtAl2002}, which makes them the key mechanism for transporting angular momentum away from the star, allowing it to grow in mass \citep[][]{UchidaShibata1985,ShuAdamsLizano1987,KoeniglPudritz2000,PudritzEtAl2007,FrankEtAl2014}. Similar to the mass transfer discussed in Section~\ref{sec:masstransfer}, we introduce a fraction $\fa$ of the accreted angular momentum $\ssink'-\ssink$ given by Equation~(\ref{eq:sinkaccretion}) and release that fraction of angular momentum along the sink particle rotation axis $\ssink'$ after accretion (denoted with a prime) by transferring it to the rotating SGS outflow and jet components, according to
\begin{equation}
\lo = \fa \left(\ssink' - \ssink\right) \cdot {\ssink'} / {|\ssink'|}.
\end{equation}
\citet{BacciottiEtAl2002} find in Hubble Space Telescope observations of the DG Tau flow that the angular momentum fraction is \mbox{$\fa=0.6$--$1.0$}, consistent with the disk wind models by \citet{PelletierPudritz1992}, who find \mbox{$\fa\sim0.7$--$1.0$}, for sub- to super-Alfv\'{e}nic accretion flows. Using numerical simulations, \citet{BanerjeePudritz2006} and \citet{HennebelleFromang2008} measured \mbox{$\fa\sim0.5$--$2$}. The time-averaged ejected angular momentum fraction is $\fa\sim0.9$ in their simulations, consistent with the observations in \citet{BacciottiEtAl2002}, so we use $\fa=0.9$ as the standard value for our SGS model. This means that $90\%$ of the accreted angular momentum is transferred to the outflow and removed from the disk--protostar system, which is a reasonable value for magnetically-driven jets and outflows \citep{PudritzEtAl2007}.

\subsection{Summary of all SGS outflow parameters} \label{sec:SGSparams}

\begin{table}
\caption{List of SGS outflow parameters.}
\label{tab:SGSparams}
\def\arraystretch{1.3}
\begin{tabular*}{\linewidth}{@{\extracolsep{\fill} }lccr}
\hline
\hline
SGS Parameter & Symbol & Default & Reference \\
\hline
Outflow Opening Angle & $\thetaout$ & $30^\circ$ & [1] \\
Mass Transfer Fraction & $\fmass$ & $0.3$ & [2] \\
Jet Speed Normalization$^{a}$ & $|\vo|$ & $100\,\km\,\s^{-1}$ & [3] \\
Angular Momentum Fraction & $\fa$ & $0.9$ & [4] \\
Outflow Radius & $\rout$ & $16\,\dx$ & Section~\ref{sec:tests} \\
\hline
\end{tabular*}
\\
\textbf{Notes.} $^a\,$The outflow velocities are dynamically computed according to the Kepler speed at the footpoint of the jet, $|\vo|=100\,\km\,\s^{-1}(\ms/0.5\,\msol)^{1/2}$ (see Equation~\ref{eq:vrad}). References: [1] \citet{BlandfordPayne1982,AppenzellerMundt1989,Camenzind1990,CasseKeppens2002}; [2] \citet{HartmannCalvet1995,Calvet1998,Tomisaka1998,BacciottiEtAl2002,Tomisaka2002,LeeEtAl2006,CabritEtAl2007,LeeEtAl2007,HennebelleFromang2008,DuffinPudritz2009,BacciottiEtAl2011,PriceTriccoBate2012,SeifriedEtAl2012,SheikhnezamiEtAl2012,FendtSheikhnezami2013}; [3] \citet{Herbig1962,SnellLorenPlambeck1980,BlandfordPayne1982,Draine1983,UchidaShibata1985,ShibataUchida1985,ShibataUchida1986,PudritzNorman1986,WardleKoenigl1993,BacciottiEtAl2000,KoeniglPudritz2000,BacciottiEtAl2002,BanerjeePudritz2006,MachidaEtAl2008}; [4] \citet{PelletierPudritz1992,BacciottiEtAl2002,BanerjeePudritz2006,HennebelleFromang2008}.
\end{table}

\begin{table*}
\caption{List of outflow test simulations.}
\label{tab:testsims}
\def\arraystretch{1.3}
\begin{tabular*}{\linewidth}{@{\extracolsep{\fill} }lrrrrrrr}
\hline
\hline
Simulation Model & SGS On/Off & $\lref$ & $\dx\;[\AU]$ & $\rsink\;[\AU]$ & $\rhosink\;[\g\,\cm^{-3}]$ & $\rout\;[\dx]$ & $\jres\;[\dx]$ \\
(1) & (2) & (3) & (4) & (5) & (6) & (7) & (8) \\
\hline
(01) \texttt{NoSG\_L10} &         Off &    $10$ &    $ 7.8$ &      $ 19.6$ &   $  1.1\times10^{-13}$ &     n/a &    $32$ \\
(02) \texttt{NoSG\_L11} &         Off &    $11$ &    $ 3.9$ &      $  9.8$ &   $  9.0\times10^{-13}$ &     n/a &    $32$ \\
(03) \texttt{NoSG\_L12} &         Off &    $12$ &    $ 2.0$ &      $  4.9$ &   $  9.1\times10^{-12}$ &     n/a &    $32$ \\
(04) \texttt{NoSG\_L13} &         Off &    $13$ &    $ 1.0$ &      $  2.4$ &   $  9.2\times10^{-11}$ &     n/a &    $32$ \\
(05) \texttt{NoSG\_L14} &         Off &    $14$ &    $ 0.5$ &      $  1.2$ &   $  9.2\times10^{-10}$ &     n/a &    $32$ \\
\hline
(06) \texttt{SGSM\_L07} &          On &    $ 7$ &    $62.7$ &      $ 156.7$ &   $  1.1\times10^{-15}$ &    $16$ &    $32$ \\
(07) \texttt{SGSM\_L08} &          On &    $ 8$ &    $31.3$ &      $ 78.3$ &   $  5.1\times10^{-15}$ &    $16$ &    $32$ \\
(08) \texttt{SGSM\_L09} &          On &    $ 9$ &    $15.7$ &      $ 39.2$ &   $  2.4\times10^{-14}$ &    $16$ &    $32$ \\
(09) \texttt{SGSM\_L10} &          On &    $10$ &    $ 7.8$ &      $ 19.6$ &   $  1.1\times10^{-13}$ &    $16$ &    $32$ \\
(10) \texttt{SGSM\_L11} &          On &    $11$ &    $ 3.9$ &      $  9.8$ &   $  9.0\times10^{-13}$ &    $16$ &    $32$ \\
\hline
(11) \texttt{SGSM\_L11\_R4} &      On &    $11$ &    $ 3.9$ &      $  9.8$ &   $  9.0\times10^{-13}$ &     $4$ &    $32$ \\
(12) \texttt{SGSM\_L11\_R8} &      On &    $11$ &    $ 3.9$ &      $  9.8$ &   $  9.0\times10^{-13}$ &     $8$ &    $32$ \\
(13) \texttt{SGSM\_L11\_R32} &     On &    $11$ &    $ 3.9$ &      $  9.8$ &   $  9.0\times10^{-13}$ &    $32$ &    $32$ \\
\hline
(14) \texttt{SGSM\_L11\_Fa0.0}$^{a}$ &      On &    $11$ &    $ 3.9$ &      $  9.8$ &   $  9.0\times10^{-13}$ &     $16$ &    $32$ \\
(15) \texttt{SGSM\_L11\_Fm0.1}$^{b}$ &      On &    $11$ &    $ 3.9$ &      $  9.8$ &   $  9.0\times10^{-13}$ &     $16$ &    $32$ \\
\hline
(16) \texttt{NoSG\_L11\_J2} &     Off &    $11$ &    $ 3.9$ &      $  9.8$ &   $  9.0\times10^{-13}$ &     n/a &     $2$ \\
(17) \texttt{NoSG\_L11\_J4} &     Off &    $11$ &    $ 3.9$ &      $  9.8$ &   $  9.0\times10^{-13}$ &     n/a &     $4$ \\
(18) \texttt{NoSG\_L11\_J8} &     Off &    $11$ &    $ 3.9$ &      $  9.8$ &   $  9.0\times10^{-13}$ &     n/a &     $8$ \\
(19) \texttt{NoSG\_L11\_J16} &    Off &    $11$ &    $ 3.9$ &      $  9.8$ &   $  9.0\times10^{-13}$ &     n/a &    $16$ \\
(20) \texttt{NoSG\_L11\_J64} &    Off &    $11$ &    $ 3.9$ &      $  9.8$ &   $  9.0\times10^{-13}$ &     n/a &    $64$ \\
\hline
\end{tabular*}
\\
\textbf{Notes.} Columns: (1) simulation name, (2) SGS model switched on/off, (3) maximum refinement level, resulting in an effective resolution of $N_\mathrm{res}=2^{\lref}$, (4) minimum cell size, (5) sink particle radius, (6) sink particle density threshold, (7) SGS outflow radius, (8) Jeans length resolution. All simulations share the same initial conditions for the star-forming core: $\rho=3.82\times10^{-18}\,\g\,\cm^{-3}$, $R=5\times10^{16}\,\cm$, $M=1\,\msol$, $\tff=1.075\times10^{12}\,\s=34\,\mathrm{k}\yr$, $\Omega=1.86\times10^{-13}\,\s^{-1}$, and $B_z=100\,\mu\Gauss$ (see text for details). \\$^{a}$Run \texttt{SGSM\_L11\_Fa0.0} is identical to the standard SGS run \texttt{SGSM\_L11}, but angular momentum transfer is switched off ($\fa=0$). \\$^{b}$Run \texttt{SGSM\_L11\_Fm0.1} is also identical to \texttt{SGSM\_L11}, but the mass transfer fraction was set to $\fmass=0.1$, instead of the standard $\fmass=0.3$.
\end{table*}

In summary, our SGS outflow model employs the user-adjustable parameters listed in Table~\ref{tab:SGSparams}. The default outflow opening angle $\thetaout=30^\circ$, the mass transfer fraction $\fmass=0.3$, the jet speed $|\vo|=100\,\km\,\s^{-1}(\ms/0.5\,\msol)^{1/2}$, and the angular momentum transfer fraction $\fa=0.9$. These default parameters are all physically motivated and chosen based on theoretical models, numerical simulations and observations. The only freely tunable parameter of our SGS model is the outflow radius. By varying $\rout$ from 4 to 32 grid cells on the highest level of AMR in the test simulations discussed in the next section, we find that all relevant outflow properties (mass, momentum, angular momentum, and jet speed) are converged to within $25\%$ for $\rout=16\,\dx$, so we define this value as the default for our SGS outflow model.

\subsection{Using magnetic fields together with the SGS model}
We briefly note that our SGS model would technically work in simulations without a magnetic field, but we do not recommend to use it in non-MHD simulations. We ran test simulations such as discussed in the next section, but with $B=0$. We find that the magnetic field already modifies the collapse and evolution of the dense core well before a sink particle forms and thus well before the SGS model kicks in. In fact, with $B=0$, the whole core and disk fragments into a few objects rather than forming only a single star in the center. It is known that magnetic fields reduce fragmentation and reduce the star formation rate \citep[e.g.,][]{PriceBate2007,HennebelleTeyssier2008,BuerzleEtAl2011,PetersEtAl2011,HennebelleEtAl2011,SeifriedEtAl2011,FederrathKlessen2012}. It is thus not surprising that we find fragmentation when we do not include a magnetic field. Given that dense cores are magnetized with mass-to-flux ratios of the order of 5 \citep{CrutcherEtAl2010} we performed all our simulations using such initial conditions.

\section{Tests of the outflow model} \label{sec:tests}

In order to test our SGS model, we compare it to simulations that \emph{do not have} the SGS outflow model, but increase the resolution step by step to gradually resolve the jet-launching mechanism. Those models produce outflows self-consistently by releasing energy from the wound-up magnetic field, similar to previous such simulations \citep[e.g.,][]{BanerjeePudritz2006,HennebelleFromang2008,MachidaEtAl2008,BuerzleEtAl2011SPHoutflows,SeifriedEtAl2012,PriceTriccoBate2012,BateTriccoPrice2014}. The problem is that the jet and outflow properties of those simulations strongly depend on the resolution, because the jet is launched from the innermost part of the accretion disk, close to the protostar, which is often not sufficiently resolved. Even if it is resolved, then the simulation cannot be run for a long time after protostar formation, as in e.g., \citet{BateTriccoPrice2014}, who can only follow their highest-resolution run for $2\,\yr$ after protostar formation, a phase in which the jet and outflow components are just starting to build up momentum. Here we present a resolution study of self-consistently launched outflows and compare them with simulations including our SGS model.

\begin{figure*}
\centerline{\includegraphics[width=1.0\linewidth]{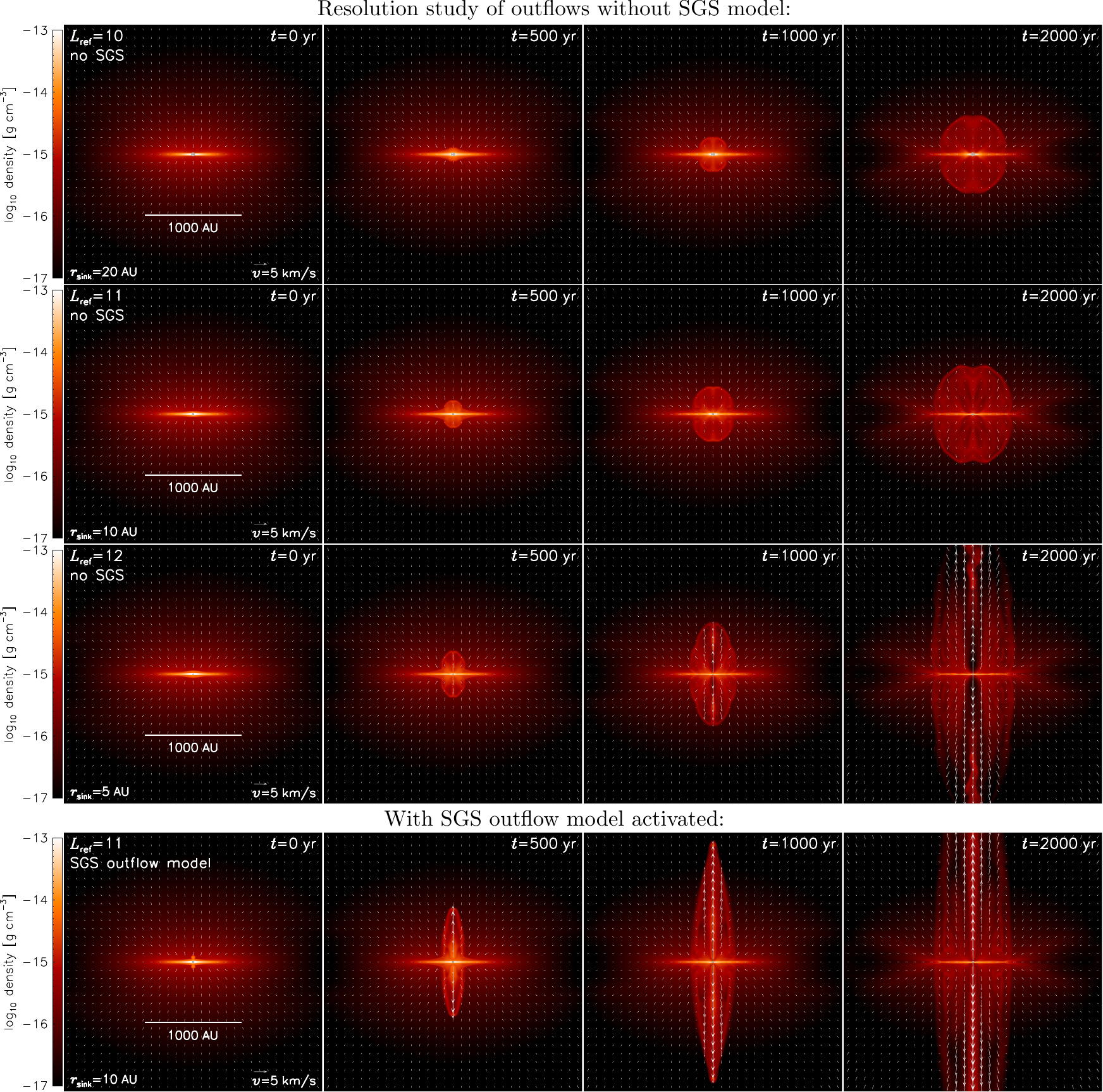}}
\caption{Snapshots of the density structure perpendicular to the disk midplane of a collapsing, rotating cloud core, forming a single star in the center and driving a bipolar outflow. Columns from left to right show different times, $t=0$, $500$, $1000$, and $2000\,\yr$ after protostar formation. The top three rows show runs \emph{without} the SGS model and with increasing grid resolution (refinement levels $\lref=10$, $11$, $12$), while the last row shows the $\lref=11$ run, but \emph{with our SGS outflow model activated} (see Table~\ref{tab:testsims} for details of the calculations). Velocity vectors normalized to $5\,\km\,\s^{-1}$ are superimposed. Simulations without the SGS model exhibit a strong resolution dependence with a qualitative change occurring between $\lref=11$ and $12$. For $\lref\leq11$, the outflow only consists of an expanding, nearly spherical bubble, while for $\lref\ge12$, a fast jet component forms, which is, however, still under-resolved. With our SGS model activated (bottom row), we recover the fast jet component at moderate resolution and obtain converged outflow properties, which would require a significantly higher resolution without SGS model (see the more detailed analyses in Figure~\ref{fig:totals_lref}). (\emph{Movies are available in the online version.})}
\label{fig:lrefimages}
\end{figure*}

\subsection{Initial conditions}

Our test simulations use the standard initial conditions for isolated disk and star formation, similar to e.g., \citet{BanerjeePudritz2006}, \citet{MachidaEtAl2008}, \citet{HennebelleFromang2008} or \citet{PriceTriccoBate2012}. All our simulations start with a uniform, spherical gas distribution with a density of $\rho=3.82\times10^{-18}\,\g\,\cm^{-3}$ and a radius of $R=5\times10^{16}\,\cm$, embedded in a computational box of side length $L=1.2\times10^{17}\,\cm=0.04\,\pc$. The mass of the core is $M=1\,\msol$. We impose an initial solid-body rotation along the $z$-axis with an angular frequency of $\Omega=1.86\times10^{-13}\,\s^{-1}$. With the mean freefall time $\tff=1.075\times10^{12}\,\s=34\,\mathrm{k}\yr$, this results in a rotation--to--gravity parameter of $\Omega\times\tff=0.2$, similar to previous studies of single-core collapse \citep{BanerjeePudritz2006,MachidaEtAl2008}. The initial magnetic field is uniform along the $z$-axis with a strength of $B_z=100\,\mu\Gauss$, corresponding to a typical initial mass--to--flux ratio of $(M/\Phi)/(M/\Phi)_\mathrm{crit}=5.2$ with the critical mass--to--flux ratio $(M/\Phi)_\mathrm{crit}=0.53/(3\pi)(5/G)^{1/2}=487\,\g\,\cm^{-2}\,\Gauss^{-1}$ \citep{MouschoviasSpitzer1976}. The initial energy ratios of the core are $E_\mathrm{rot}/E_\mathrm{grav}=0.022$ and $E_\mathrm{mag}/E_\mathrm{grav}=0.065$. The Jeans length is always resolved with 32 grid cells, except for the Jeans resolution study in Appendix~\ref{app:jeans}, where we show that using 32 cells per Jeans length yields converged results.

In the following, we vary the resolution, i.e., the maximum refinement level $\lref$ for cases without and with SGS model, and we vary the outflow radius $\rout$, in order to determine the impact of changing the resolution on the outflow morphology, the mass, momentum, angular momentum and maximum speed of the outflow. Table~\ref{tab:testsims} provides a list of all the test simulations and their characteristic parameters.

\subsection{Outflow morphology}

In order to get a basic overview of the time evolution and morphology of the outflows in the test simulations, we present snapshots of the density structure in Figure~\ref{fig:lrefimages}. During the collapse of the rotating core, a disk forms with a protostar in the center. The columns in Figure~\ref{fig:lrefimages} show slices of the gas density perpendicular to the disk midplane when $t=0$, $500$, $1000$, and $2000\,\yr$ after protostar formation. The first three rows of Figure~\ref{fig:lrefimages} show simulations \texttt{NoSG\_L10}, \texttt{NoSG\_L11}, and \texttt{NoSG\_L12}, i.e., simulations \emph{without} the SGS outflow model and increasing maximum refinement level $\lref=10$, $11$, and $12$. We see an outflow forming in all of them, but the outflow morphology and speed exhibit a strong dependence on the numerical resolution. Model \texttt{NoSG\_L10} produces a roughly spherical expansion wave, while model \texttt{NoSG\_L12} clearly shows a bipolar outflow along the rotation axis of the disk. Comparing models \texttt{NoSG\_L11} and \texttt{NoSG\_L12}, we not only see a quantitative change in behavior, but also a qualitative change. In model \texttt{NoSG\_L12}, we can clearly distinguish a collimated high-speed jet and a broad, low-speed outflow, while the jet component is basically absent in models \texttt{NoSG\_L10} and \texttt{NoSG\_L11}. We emphasize though that even run \texttt{NoSG\_L12} is not yet converged with resolution.

Inspection of all the test simulations shows that models without the SGS outflow model and with $\lref\geq12$, corresponding to a physical resolution of $\dx\leq2\,\AU$ (see Table~\ref{tab:testsims}) have an outflow+jet component, while simulations with $\lref<12$ only resolve the low-speed outflow component and yield no high-speed jet. However, even if $\lref\geq12$ and the jet component is present, its mass, momentum, and speed depend on the numerical resolution. Convergence is expected when the launching radius close to the protostar is actually resolved, which requires a computationally-prohibitive refinement level of $\lref\sim17$, as estimated and discussed in detail in the next section. Thus, the density and velocity structure of the jet and outflow in model \texttt{NoSG\_L12} (3rd row in Figure~\ref{fig:lrefimages}) is not quantitatively correct, because it is not fully resolved.

The bottom row of Figure~\ref{fig:lrefimages} shows the same simulation as in the 2nd row ($\lref=11$), but $\emph{with}$ our SGS outflow model activated. We now recover the high-speed jet component, which was absent without SGS model at that resolution (compare to 2nd row), and we find converged results for all relevant global outflow quantities when our SGS outflow model is activated, which we demonstrate next.

\subsection{Outflow mass, momenta and speeds}

\subsubsection{Resolution effects in simulations without SGS model}

\begin{figure*}
\centerline{\includegraphics[width=0.92\linewidth]{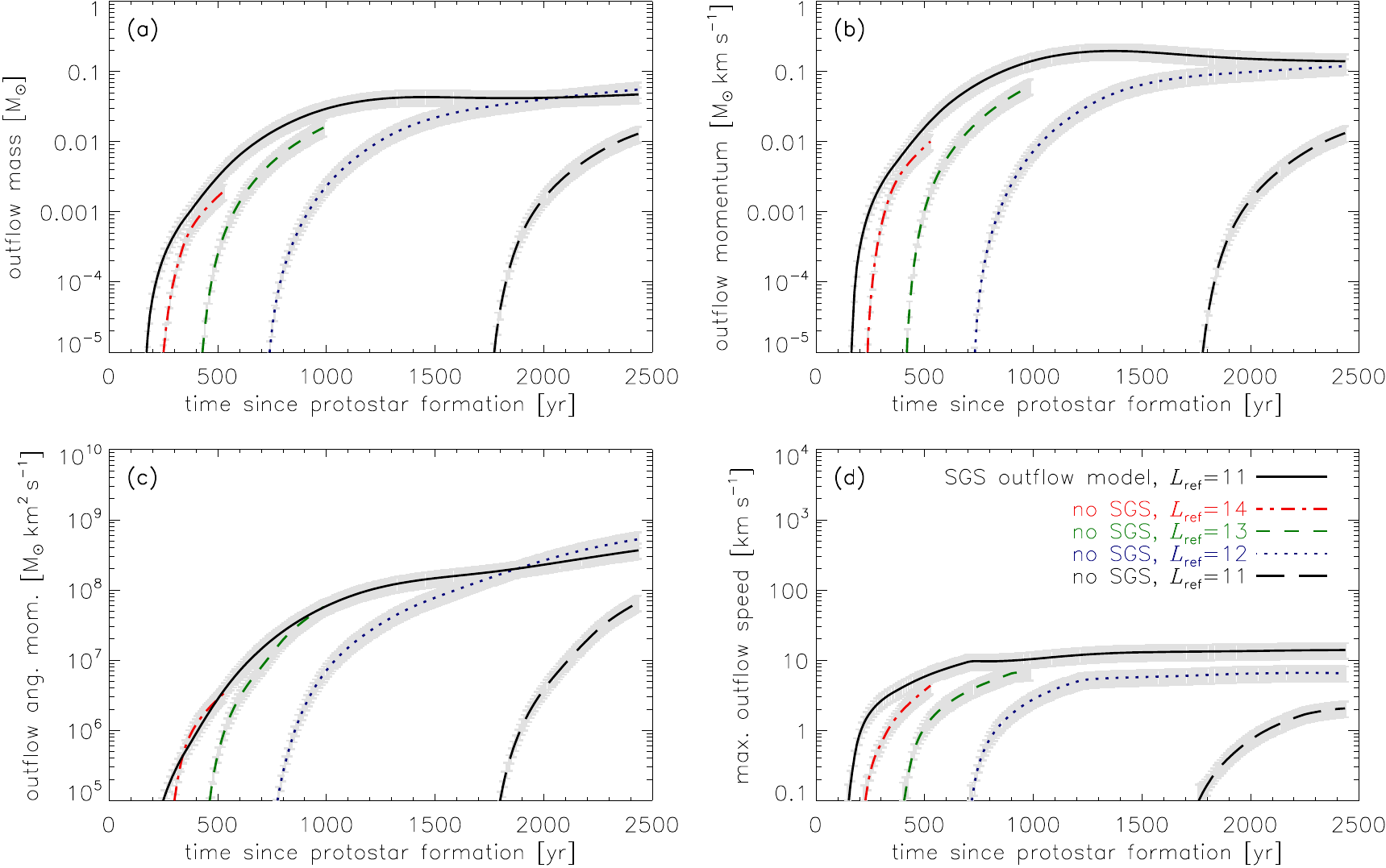}}
\caption{Time evolution of the outflow mass (a), momentum (b), angular momentum (c), and maximum outflow speed (d), in test simulations of isolated disk and star formation. All outflow properties were measured within two cylindrical volumes with radius and height of $500\,\AU$, located $500\,\AU$ above or below the disk, respectively. The solid line shows the result of our SGS model for a refinement level $\lref=11$ (test model \texttt{SGSM\_L11} in Table~\ref{tab:testsims}), while the other curves show models without SGS model, but increasing resolution, corresponding to refinement levels $\lref=11$--$14$ (see Table~\ref{tab:testsims} for details). Without the SGS model, the outflow mass, momentum, angular momentum and speed depend on the resolution of the simulation. Near convergence is reached with $\lref\gtrsim14$, corresponding to a physical resolution of $\dx\lesssim0.5\,\AU$, while full convergence requires $\lref\sim17$, in order to resolve the typical launching radius of $10\,\rsol$. With the SGS model, we find outflow masses, momenta and speeds that agree within a factor of two of the converged values at much lower resolution. The gray error bars on each curve indicate 25\% uncertainty intervals.}
\label{fig:totals_lref}
\end{figure*}

Figure~\ref{fig:lrefimages} demonstrated qualitatively that models \emph{without} the SGS model exhibit a strong resolution dependence. Here we investigate this in detail, by quantifying the mass, momentum, angular momentum and speed of the outflow for all the test simulations in Table~\ref{tab:testsims}. Figure~\ref{fig:totals_lref} shows the time evolution of the mass (panel a), momentum (panel b), angular momentum (panel c), and maximum speed (panel d) of the outflow in simulations \emph{without} the SGS model and with increasing refinement level $\lref=11$--$14$. We also show our standard simulation with the SGS outflow model activated for an intermediate refinement level $\lref=11$ (solid line).

We measure the outflow properties (mass, momentum, angular momentum, and speed) at each time, by selecting all computational cells within two cylindrical volumes above and below the disk with an outflowing vertical velocity component, $v_z>0$ for $z>0$ or $v_z<0$ for $z<0$, where $z$ denotes the rotation axis of the disk and the sink particle. The radius and height of the two cylinders is $500\,\AU$ and they are located respectively $500\,\AU$ above and below the disk, with their symmetry axes centered on the rotation axis of the disk. We chose this configuration, because it excludes cells inside the SGS volume for any of our test simulations (except for run \texttt{SGSM\_L07}, which has such a low resolution that the outflow radius $\rout\sim1000\,\AU$ reaches into the analysis volume). This choice of analysis volume is a good compromise, because of two reasons. First, we want to get the large-scale outflow properties right with the SGS model, so we measure them $500\,\AU$ above and below the disk, and we do not want to include the SGS volume in the analysis to avoid any bias from cells that were directly affected by the SGS model. Second, most of the runs without SGS outflow model produce outflows that are far too slow compared to the converged solution. To enable a meaningful comparison, those outflows must be able to reach the analysis volume, which is why we chose this geometry for the analysis: sufficiently far away from the disk, but still close enough for the un-converged outflows to reach it.

We also made the same analysis by selecting all cells with a vertical position of one scale height $H\sim\cs\Omega^{-1}\sim\cs v_\phi^{-1}R = 25\,\AU$ above and below the disk midplane, $\left|z\right| > H$, which yields very similar relative outflow properties between the different models; only the total amount of outflowing mass, momentum, etc., is shifted, because the analysis volume is larger in this case. We further experimented with other choices for selecting the outflowing material, e.g., $\left|z\right| > 2H$ and ${\bf v}\cdot{\bf r} > 0$, where ${\bf r}$ is the position vector with respect to the center of the disk, and again obtained very similar results, differing by only a few percent from our standard selection criterion.

\begin{figure*}
\centerline{\includegraphics[width=0.92\linewidth]{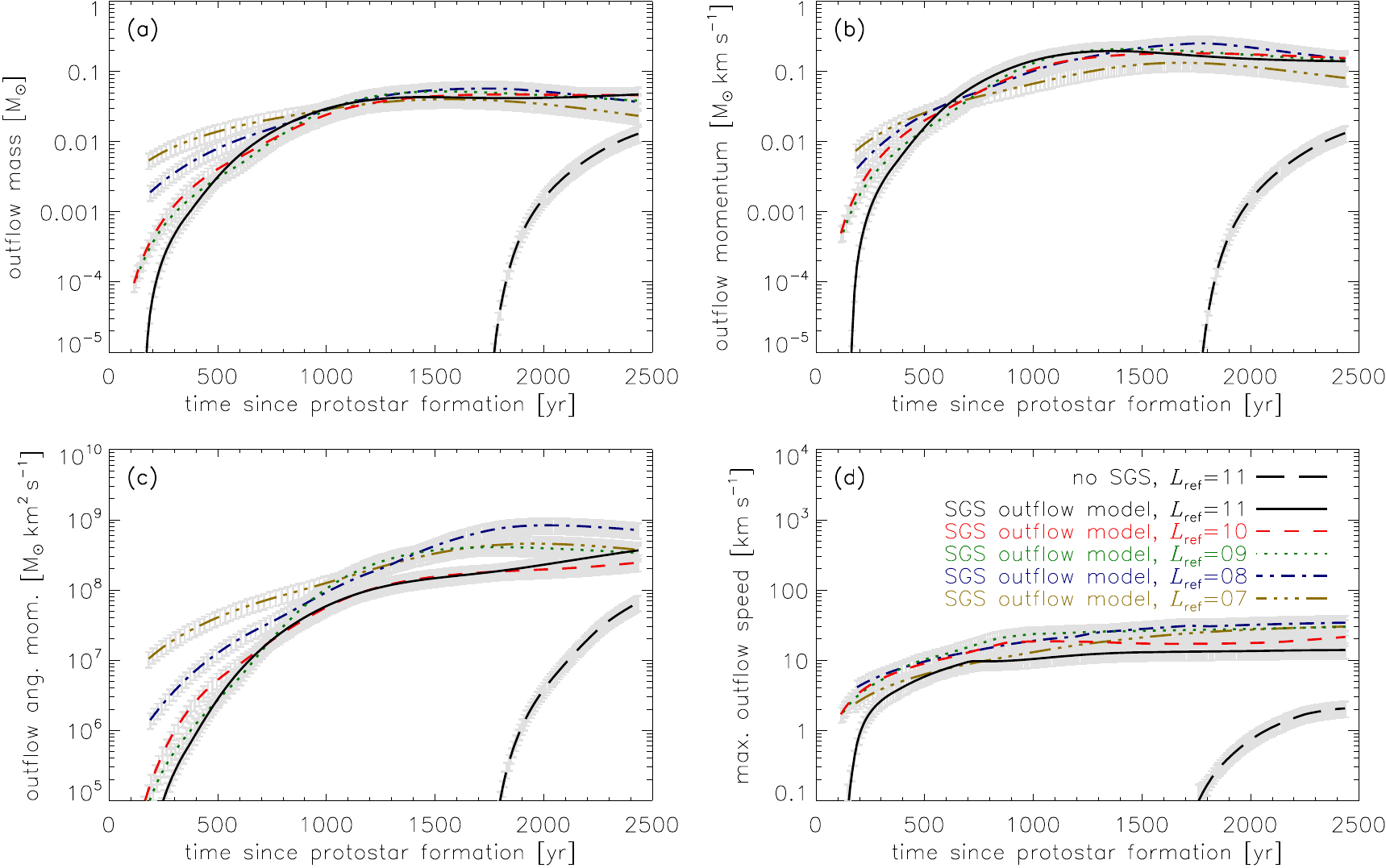}}
\caption{Same as Figure~\ref{fig:totals_lref}, but for calculations with SGS model activated and different grid resolutions, $\lref=7$--$11$. Although the resolution is varied by a factor of 16 and the lowest-resolution run (\texttt{SGSM\_L07}) with $\dx\sim60\,\AU$ has a large outflow radius of $\rout\sim1000\,\AU$, we find that all outflow properties are converged to within a factor of two at late times. In contrast, the long-dashed line shows run \texttt{NoSG\_L11} (without SGS model and $\lref=11$), which is clearly under-resolved and does not produce any of the converged outflow properties.}
\label{fig:totals_sgs}
\end{figure*}

\begin{figure*}
\centerline{\includegraphics[width=0.92\linewidth]{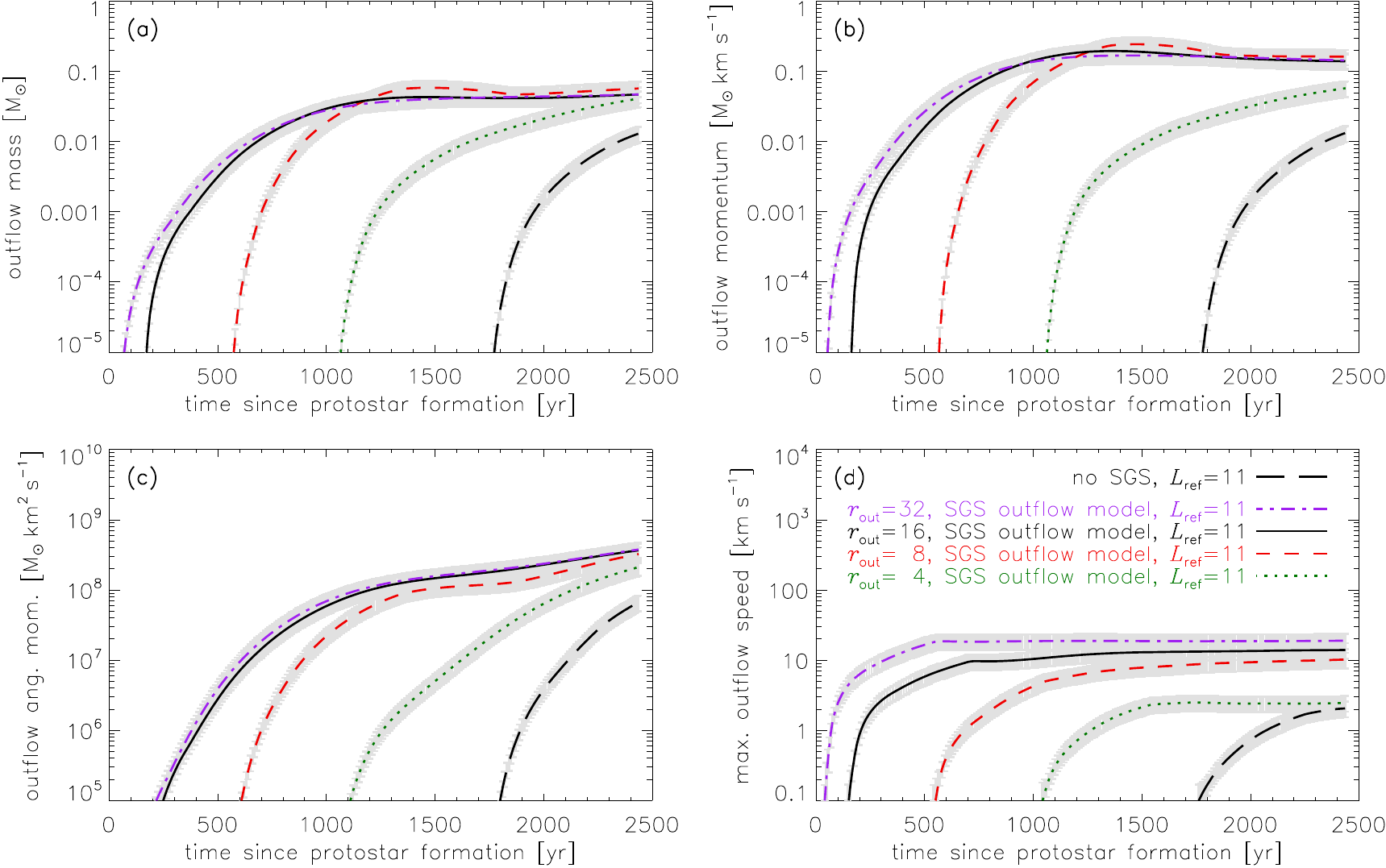}}
\caption{Same as Figure~\ref{fig:totals_sgs}, but for calculations with different SGS outflow radii, $\rout/\dx=4$, $8$, $16$, $32$. We find convergence of the outflow mass, momentum, angular momentum and maximum outflow speed to within $25\%$, if $\rout\ge16\,\dx$. The run with $\rout=8\,\dx$ underestimates the jet speed by $\sim50\%$, but is otherwise well converged at late times. The run with $\rout=4\,\dx$ transfers insufficient linear and angular momentum, and the jet speed is about an order of magnitude too low.}
\label{fig:totals_rout}
\end{figure*}

\begin{figure*}
\centerline{\includegraphics[width=0.92\linewidth]{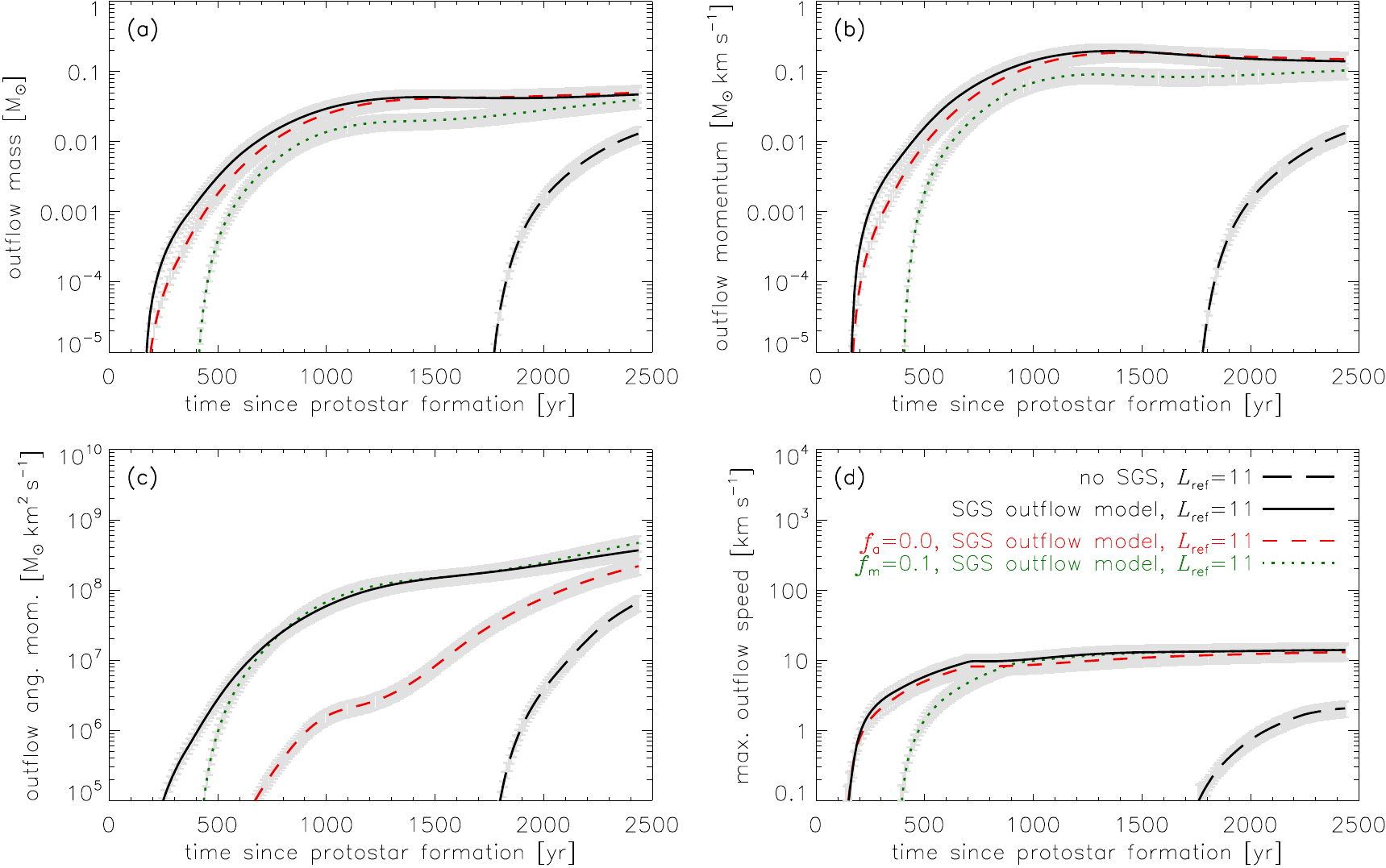}}
\caption{Same as Figure~\ref{fig:totals_rout}, but for runs where the SGS parameter $\fa=0.0$, i.e., no angular momentum transfer (standard is $\fa=0.9$) or $\fmass=0.1$, i.e., only 10\% mass transfer (standard is $\fmass=0.3$). Comparing to our standard SGS run \texttt{SGSM\_L11}, we find that switching off angular momentum transfer yields results identical to the standard case, except for a clear deficit in angular momentum of the outflowing gas. Using a mass transfer fraction $\fmass=0.1$ yields identical results to $\fmass=0.3$, but with an outflow mass and momentum about $15\%$ smaller at late times. This relatively small reduction is due to self-regulation of outflow feedback.}
\label{fig:totals_misc}
\end{figure*}

Figure~\ref{fig:totals_lref} shows that the outflow properties in simulations without the SGS outflow model slowly converge with resolution. Even our highest-resolution model with a refinement level of $\lref=14$ and a sub-AU resolution of $\dx=0.5\,\AU$ is not fully converged and we could only afford to run it until about $500\,\yr$ after protostar formation. This is still significantly longer than the state-of-the-art, fully-resolved smoothed particle magnetohydrodynamics (SPH) simulations by \citet{BateTriccoPrice2014}, which had to be stopped at $t=2\,\yr$, clearly too early to yield a fully-developed outflow and jet. The black solid lines in Figure~\ref{fig:totals_lref} show that our standard simulation with SGS outflow model activated (\texttt{SGSM\_L11}) yields an outflow mass, momentum, angular momentum and speed comparable to or slightly higher than our highest-resolution simulation without SGS model (\texttt{NoSG\_L14}). These values are all converged to within a factor of two of the extrapolated, fully-converged values expected for $\lref\sim17$. Thus, our SGS model produces outflow properties at moderate resolutions that would otherwise require significantly higher resolution.

We note that these outflow properties are similar to, but somewhat smaller than typical outflow parameters measured in observations by \citet{MauryAndreLi2009}, \citet{ArceEtAl2010}, \citet{CurtisEtAl2010}, \citet{GinsburgBallyWilliams2011}, \citet{PlunkettEtAl2013}, and more recently in the P6-SMA6 core in the Galactic `Snake' infrared dark cloud G11.11-0.12 by \citet{WangEtAl2014Henning}, who all find outflow masses of \mbox{$\sim0.1$--$1\,\msol$}, outflow momenta of \mbox{$\sim0.5$--8$\,\msol\,\km\,\s^{-1}$}, and speeds of $\sim30\,\km\,\s^{-1}$ for a dynamical age of $\sim10^4\,\yr$. The differences arise because our test simulations only follow the outflows for a few $\times10^3\,\yr$, and because we measure all outflow properties in a relatively small volume, while observations usually probe larger volumes. Indeed, when we select \emph{all} cells with an outflow velocity above and below the disk, we find values in the observed ranges for the mass and momentum. We also note that the outflow speeds of $\sim10$--$20\,\km\,\s^{-1}$ are similar to observations by \citet{RagaEtAl2013} and smaller than the normalization speed (Kepler speed) at the footpoint of the jet ($100\,\km\,\s^{-1}$), because the high-speed jet shocks and is decelerated by the ambient gas, leading to significant entrainment. Using a passive scalar tracer of the material that has been accreted and was subsequently re-injected by our SGS outflow model, we measure that about $40\%$ of the outflowing gas is launched directly in the jet and the remaining $60\%$ of the outflowing gas is entrained material from the collapsing core envelope.

\subsubsection{Convergence study of the SGS outflow model} \label{sec:SGSconvergence}

\paragraph{Dependence on the level of refinement}
A reasonable SGS model is expected to yield results that are similar to an otherwise equivalent calculation without SGS model carried out at higher resolution. This is demonstrated in our preceding discussion of Figures~\ref{fig:lrefimages} and~\ref{fig:totals_lref}. However, a properly working SGS model is also expected to be adaptive and yield converged results, even if the resolution of the simulation is varied. We test this with simulations \texttt{SGSM\_L07}, \texttt{SGSM\_L08}, \texttt{SGSM\_L09}, \texttt{SGSM\_L10} and \texttt{SGSM\_L11}, shown in Figure~\ref{fig:totals_sgs}. These simulations are identical to all the previous ones, expect that we now vary the refinement level $\lref=7$--$11$, while our SGS outflow model is activated. Figure~\ref{fig:totals_sgs} shows that all outflow properties are converged at late times to within a factor of two for different resolutions, in contrast to the strong resolution dependence seen in simulations without SGS model (cf.~Figure~\ref{fig:totals_lref}).

\paragraph{Dependence on the number of cells per outflow radius}
As discussed in Section~\ref{sec:SGSparams} and listed in Table~\ref{tab:SGSparams}, our SGS outflow model has five basic parameters, with the first four of them being determined and fixed by observations, theoretical models and numerical simulations. The only tunable parameter is the SGS outflow radius $\rout$, which is of purely numerical nature and determines the size (volume) of the outflow launching region (see Figure~\ref{fig:schematic}). All the outflow injection properties are formulated and implemented in Section~\ref{sec:SGSmodel} such that the total mass, momentum and angular momentum transfer do not depend on the volume of the SGS launching region. Due to the discretization of the computational grid, however, the choice of the number of cells within the SGS launching region will have a profound consequence for how well the jet and outflow components will actually transfer mass, momentum and angular momentum and what maximum jet speeds can be reached. In order to find a minimum number of cells required for convergence of our SGS model, we now vary $\rout$ in units of $\dx$, while keeping everything else fixed.

Figure~\ref{fig:totals_rout} shows a comparison of runs \texttt{SGSM\_L11\_R4}, \texttt{SGSM\_L11\_R8}, \texttt{SGSM\_L11} and \texttt{SGSM\_L11\_R32} with $\rout/\dx=4$, $8$, $16$, and $32$, respectively. With $\rout=4\,\dx$, all outflow properties except the mass, are significantly underestimated. For $\rout\ge8\,\dx$, we find that the mass, linear and angular momentum of the outflow are well converged at late times. Only the maximum jet speed is $\sim50\%$ too small with $\rout=8\,\dx$, while it is converged to within $25\%$ for $\rout=16\,\dx$, compared to $\rout=32\,\dx$. Given the relatively small difference for the maximum jet speed and given that the mass, momentum and angular momentum transfer are converged to within $\lesssim5\%$ for $\rout=16\,\dx$, we define $\rout=16\,\dx$ as the standard choice for the SGS outflow radius\footnote{Setting $\rout=16\,\dx$ instead of the fully converged value ($\rout=32\,\dx$) is a compromise based on the fact that a larger SGS outflow volume requires significantly more computational time to loop over (with our standard choice, $\rout=16\,\dx$, we have about 1,150 AMR cells per outflow cone, while with $\rout=32\,\dx$, there are about 9,200 cells, i.e., a factor of eight more, as expected for a three-dimensional calculation). A larger SGS volume also requires significantly more communication, when the SGS outflow regions extend across more than one processor. We thus find the $25\%$ difference in jet speeds between the standard choice, $\rout=16\,\dx$, and the fully converged case a reasonable compromise.} (see Table~\ref{tab:SGSparams}). We note that this corresponds to a diameter of the SGS region of $32\,\dx$, which agrees well with the required Jeans resolution of $32\,\dx$ for convergence (see Appendix~\ref{app:jeans}).

\paragraph{Dependence on the angular momentum and mass transfer fractions}
It is also interesting to investigate how our results depend on whether angular momentum transfer is included or not. Similarly, some theoretical models favor a mass transfer fraction of $\fmass=0.1$ instead of $\fmass=0.3$, as we have discussed in Section~\ref{sec:masstransfer}. Figure~\ref{fig:totals_misc} shows the effect of removing the angular momentum transfer from the SGS model, by setting the angular momentum transfer fraction to $\fa=0.0$. We see that everything remains the same, except for the angular momentum in the outflowing material, which is $\sim60\%$ too small. The angular momentum transfer feature of our SGS model thus seems to be a significant improvement over previous SGS outflow implementations, which all neglect angular momentum transfer (see discussion in Section~\ref{sec:discussion}).

Figure~\ref{fig:totals_misc} also shows a model with the mass transfer fraction set to $\fmass=0.1$. The mass and linear momentum of the outflowing gas resulting from this model are smaller by only $\sim15\%$ at late times, compared to our standard case with $\fmass=0.3$. This may seem surprising at first, because now only 10\% of the accreted material is launched as an outflow/jet compared to 30\% inside the SGS region and one might have guessed that the ejected mass should be smaller by a factor of three rather than by only 15\%. However, what we see here is a manifestation of self-regulation by feedback. Ejecting only $\fmass=10\%$ of the accreted material per timestep means that more gas is accreted compared to when $\fmass=30\%$. But this leads to a higher \emph{absolute} accretion rate in each time step, which in turn leads to a higher absolute outflow rate, thus nearly compensating the effect of lowering the $\fmass$ fraction and finally leading to a very similar total outflow rate as with $\fmass=30\%$. Thus, as long as $\fmass$ is in the reasonable range of $0.1$--$0.3$, the SGS mass transfer fraction has only little effect on the outflow properties outside the launching region. For instance, \citet{PetersEtAl2014} used a preliminary version of our SGS outflow model with $\fmass=0.1$ and applied it to massive star formation. Our tests here show that $\fmass=0.1$ is a suitable choice and only leads to a minor underestimate of the mass and momentum injected by the outflow.

\paragraph{Properties inside the SGS launching region and saving in computational time}
Finally, we note the caveat that even though the large-scale mass, momentum, angular momentum transfer and jet speeds are well converged with our SGS outflow model, the internal, small-scale morphology of the jet and outflow still depend on the numerical resolution. It is clear that structures below a given maximum resolution cannot be resolved. Using our SGS model, however, we recover the total mass, momentum, energy, and angular momentum injection, as well as the jet speeds exerted onto super-resolution scales. This allows us to obtain global outflow properties that are indeed converged within a factor of two of the fully-resolved limit, even with resolutions that are a factor of several hundred times smaller than required without the SGS outflow model. For the particular cases studied here, we demonstrated convergence to within a factor of two for a refinement level of $\lref=7$ ($\dx\sim63\,\AU$; see Table~\ref{tab:testsims}) with the SGS model activated. In contrast, one requires $\lref\sim17$ without the SGS model for convergence, which means that we can afford a factor of $\sim1,\!000$ times lower resolution when the SGS outflow module is used. We then still obtain the mass, momentum and angular momentum transfer equivalent to a fully-resolved calculation without SGS model, which would take several orders of magnitude more computational time\footnote{A rough estimate by comparing runs \texttt{NoSG\_L14} and \texttt{SGSM\_L07} yields a saving in computational time by a factor of $\sim1,000$, but we only reached $\lref=14$. Running a fully-converged calculation with $\lref=17$ would at least require another factor of $8$ more computational time (considering the CFL criterion only and assuming that the total number of cells does not increase significantly), so using the SGS model with a maximum resolution of $\dx\sim60\,\AU$ roughly saves a factor of 1,000--10,000 in computational time.}. This enormous saving in resolution allows us to include the effect of outflows and jets from newborn stars during the star cluster formation calculations presented in Section~\ref{sec:cluster} below, which would otherwise be impossible with currently available computer technology.

\section{Comparison with previous SGS outflow implementations} \label{sec:discussion}

The first study including an SGS outflow model in the context of star cluster formation was presented by \citet{LiNakamura2006} and \citet{NakamuraLi2007}. Whenever the density in a cell exceeded 100 times the mean density in their simulation, a Lagrangian particle was created and 20\% of the mass in a cubic region around that cell (with a total of $3^3=27$ grid cells) was transferred to the central particle. No accretion was implemented in their model and thus, the outflow properties were not determined by the accretion rate, but only once at the time of particle creation and the outflows were not continuously driven. While \citet{LiNakamura2006} only implemented isotropic, radial point explosions, \citet{NakamuraLi2007} added a collimated component with an outflow angle of $\thetaout=30^\circ$. The direction of the outflow axis was determined by the local magnetic field vector in their SGS models, while our outflow axis is given by the spin axis of the sink particle. The underlying idea is that jets and outflows are driven out perpendicular to a rotating disk, along the rotation axis. In contrast, by using the local magnetic field direction, one might accidentally pick up the wound-up component of the magnetic field in the disk, rather than the polar component above and below the disk. Indeed, the magneto-centrifugal acceleration mechanism \citep{BlandfordPayne1982} requires that the local magnetic field vector makes an angle of $\geq30^\circ$ with the rotation axis. Thus, the local magnetic field direction is different from the global outflow axis and will not yield the correct outflow direction in the SGS model. Neither \citet{LiNakamura2006} nor \citet{NakamuraLi2007} considered angular momentum transfer by the outflow, while our SGS outflow model transfers a calibrated amount of accreted angular momentum to the outflowing gas (see Section~\ref{sec:angmomtransfer} and Figure~\ref{fig:totals_misc}).

\citet{WangEtAl2010} and \citet{NakamuraLi2011} are improved follow-up studies of \citet{NakamuraLi2007}, which both include accretion onto sink particles. However, the outflow axis was still chosen along the local magnetic field direction as in \citet{NakamuraLi2007}, subject to the same uncertainties as discussed above. \citet{WangEtAl2010} also do not adaptively change the outflow axis, but keep it fixed throughout the simulation. This is hard to justify in a turbulent medium where the rotation axis (and the local magnetic field direction) can continuously change in space and time, because of accretion from a turbulent gas reservoir \citep{JappsenKlessen2004}. While \citet{NakamuraLi2011} use a jet speed normalization of $100\,\km\,\s^{-1}$ as in our SGS model (see Section~\ref{sec:momtransfer}), \citet{WangEtAl2010} have a significantly lower jet speed normalization of only $23\,\km\,\s^{-1}$. The outflow momentum is deposited within $\rout\leq5\,\dx$ in \citet{WangEtAl2010}. Our tests in Figure~\ref{fig:totals_rout} indicate that at least $\rout\sim16\,\dx$ is required to obtain converged outflow properties. Neither \citet{WangEtAl2010} nor \citet{NakamuraLi2011} include angular momentum transfer in their SGS models.

\citet{DaleBonnell2008} studied isotropic and collimated winds in the context of massive star formation with SPH. They compare winds driven by injecting low-mass particles and momentum-driven winds with a Monte-Carlo approach. Their SGS model---although could be made adaptive---is not used in an adaptive fashion, i.e., they do not consider a dependence of the wind properties on the accretion rate. \citet{DaleBonnell2008} turn on their winds by hand and give them an outflow rate. In their application to massive star clusters, the winds were turned on when $\sim40\%$ of the gas in their cloud was already accreted onto sink particles. In contrast, a self-consistent inclusion of outflows would have had an important effect on the global evolution at much earlier times, when the first sink particles formed in their simulations.

The currently most sophisticated SGS outflow model is presented in \citet{CunninghamEtAl2011} and applied in \citet{OffnerEtAl2011,OffnerEtAl2012}, \citet{KrumholzKleinMcKee2012}, \citet{HansenEtAl2012}, \citet{OffnerArce2014}, and \citet{MyersEtAl2014}. The outflow is launched along the rotation axis of the sink particles and the outflow mass and momentum is adapted according to the sink particle accretion rate, similar to our SGS model (see Sections~\ref{sec:masstransfer} and~\ref{sec:momtransfer}). The mass and momentum are transferred to the gas in a region with $4\leq\rout/\dx\leq8$. Tests with our own model (see Figure~\ref{fig:totals_rout}) suggest that $\rout\leq8\,\dx$ is sufficient to yield the correct mass and momentum injection, but underestimates the converged maximum outflow speeds by $\sim50\%$. Thus, the impact of the outflows may have been underestimated in previous studies using the SGS outflow model by \citet{CunninghamEtAl2011}. Our SGS model uses a default value of $\rout=16\,\dx$, which yields jet speeds converged to within $25\,\%$, significantly closer to the infinite-resolution limit. Moreover, the SGS model by \citet{CunninghamEtAl2011} does not include angular momentum transfer. Convergence properties of their SGS model were not explored.

In summary, previous SGS outflow models were becoming increasingly sophisticated over time, but none of the previous models includes angular momentum transfer. Outflows and jets, however, do rotate \citep[e.g.,][]{BacciottiEtAl2002}, as a consequence of their physical driving mechanism. Considering that angular momentum transfer is highly efficient, with $\sim90\%$ of the accreted angular momentum carried away by the outflow and jet (see Section~\ref{sec:angmomtransfer}), and considering that this is likely the main mechanism responsible for angular momentum transport away from disk \citep[e.g.,][]{ShuAdamsLizano1987,KoeniglPudritz2000,PudritzEtAl2007,FrankEtAl2014} leading to the relatively slow rotation rates of young stars \citep[][]{HartmannEtAl1986}, our SGS outflow model constitutes a significant advancement. Moreover, none of the previous SGS outflow models was tested against calculations of magnetized protostellar collapse and disk evolution without SGS model, and convergence properties were not explored. In Section~\ref{sec:tests}, we presented the first rigorous resolution and convergence study and demonstrated that our SGS outflow model does indeed transfer a converged mass, momentum, and angular momentum, and yields realistic jet speeds, at 1,000 times lower resolution than would be required without the SGS model.

\section{Star cluster formation with outflow feedback} \label{sec:cluster}
Almost all stars form in clusters \citep{LadaLada2003} and many young stellar clusters are disrupted by gas loss after their birth \citep[for dynamical N-body calculations assuming a smoothly evolving spherical background potential to mimic gas expulsion, see, e.g.,][]{Tutukov1978,KroupaAarsethHurley2001,KroupaBouvier2003,MarksKroupa2012,BanerjeeKroupa2013,BanerjeeKroupa2014}. Here we model the actual conversion of gas into stars and apply our new SGS outflow feedback model to star cluster formation in a turbulent, magnetized gas cloud, in order to study gas expulsion by jets and outflows and their impact on the SFR and IMF. We compare two calculations, one without the outflow model, which serves as a control run, and one where we include the full outflow model with the standard parameters described in Section~\ref{sec:SGSparams}. The aim is to determine the impact of outflow feedback on the dynamics and star formation during the collapse of an isolated, magnetized, turbulent cloud.

\begin{table*}
\caption{List of outflow star cluster simulations.}
\label{tab:clustersims}
\def\arraystretch{1.3}
\begin{tabular*}{\linewidth}{@{\extracolsep{\fill} }lrrrrrr}
\hline
\hline
Simulation Model & SGS On/Off & $N_\mathrm{res}^3$ & $\dx\;[\AU]$ & $\rsink\;[\AU]$ & $\rout\;[\AU]$ & $\rhosink\;[\g\,\cm^{-3}]$ \\
(1) & (2) & (3) & (4) & (5) & (6) & (7) \\
\hline
(01) \texttt{Cluster\_256\_NoSG} & Off & $256^3$ & $9.67\times10^2$ & $2.42\times10^3$ & n/a & $3.60\times10^{-18}$ \\
(02) \texttt{Cluster\_256\_SGSM} & On & $256^3$ & $9.67\times10^2$ & $2.42\times10^3$ & $1.55\times10^4$ & $3.60\times10^{-18}$ \\
(03) \texttt{Cluster\_512\_NoSG} & Off & $512^3$ & $4.83\times10^2$ & $1.21\times10^3$ & n/a & $1.44\times10^{-17}$ \\
(04) \texttt{Cluster\_512\_SGSM} & On & $512^3$ & $4.83\times10^2$ & $1.21\times10^3$ & $7.73\times10^3$ & $1.44\times10^{-17}$ \\
(05) \texttt{Cluster\_1024\_NoSG} & Off & $1024^3$ & $2.42\times10^2$ & $6.04\times10^2$ & n/a & $5.77\times10^{-17}$ \\
(06) \texttt{Cluster\_1024\_SGSM} & On & $1024^3$ & $2.42\times10^2$ & $6.04\times10^2$ & $3.87\times10^3$ & $5.77\times10^{-17}$ \\
(07) \texttt{Cluster\_2048\_NoSG} & Off & $2048^3$ & $1.21\times10^2$ & $3.02\times10^2$ & n/a & $2.31\times10^{-16}$ \\
(08) \texttt{Cluster\_2048\_SGSM} & On & $2048^3$ & $1.21\times10^2$ & $3.02\times10^2$ & $1.93\times10^3$ & $2.31\times10^{-16}$ \\
(09) \texttt{Cluster\_4096\_NoSG} & Off & $4096^3$ & $6.04\times10^1$ & $1.51\times10^2$ & n/a & $1.19\times10^{-15}$ \\
(10) \texttt{Cluster\_4096\_SGSM} & On & $4096^3$ & $6.04\times10^1$ & $1.51\times10^2$ & $9.67\times10^2$ & $1.19\times10^{-15}$ \\
\hline
\end{tabular*}
\\
\textbf{Notes.} Columns: (1) simulation name, (2) SGS outflow model switched on/off, (3) maximum effective resolution, (4) minimum cell size, (5) sink particle radius, (6) SGS outflow radius, (7) sink density threshold. Initial conditions for all star cluster simulations: $\rho=6.5\times10^{-20}\,\g\,\cm^{-3}$, $R=0.5\,\pc$, $M=500\,\msol$, $\tff=0.26\,\mathrm{M}\yr$, $\sigma_v=1\,\km\,\s^{-1}$, and $B_z=50\,\mu\Gauss$ (for details, see Section~\ref{sec:cluster_ics}).
\end{table*}

\subsection{Setup and initial conditions} \label{sec:cluster_ics}

As a guide for the initial conditions of the following simulations, we take observations of typical cluster-forming regions, sometimes referred to as molecular `clumps' inside larger molecular clouds \citep[e.g.,][]{RaganEtAl2012,NakamuraLi2014}. For the sake of simplicity, reproducibility and consistency with previous simulations of star cluster formation, we start with a spherical, homogenous clump with a typical density $\rho_0=6.5\times10^{-20}\,\g\,\cm^{-3}$ and diameter $2R=1\,\pc$, resulting in a total mass $M_\mathrm{clump}=500\,\msol$. Our choice of the mass and radius of the clump is consistent with the surveys of star-forming molecular clumps by \citet{ShirleyEtAl2003}, \citet{FaundezEtAl2004}, and \citet{FontaniEtAl2005}, compiled in \citet{FallKrumholzMatzner2010}.

Given the initial density, the freefall time of the molecular clump is
\begin{equation} \label{eq:tff}
\tff=\sqrt{3\pi/(32G\rho_0)}=0.26\,\mathrm{M}\yr\,,
\end{equation}
which we use as the basic time unit when we describe the evolution of the cloud below. The clump is initially cold with a temperature of $T_0=11\,\mathrm{K}$, corresponding to a sound speed of $\cs=0.2\,\km\,\s^{-1}$, because such clumps can cool very efficiently \citep[e.g.,][]{WolfireEtAl1995,Krumholz2014}. The equation of state follows the simple polytropic form given in Equations~(\ref{eq:eos1}) and~(\ref{eq:eos2}) to roughly describe the thermal evolution during the collapse \citep{MasunagaInutsuka2000}. In order to establish initial pressure equilibrium, the clump is embedded in a warm, diffuse medium with density $10^{-2}\,\rho_0$ and correspondingly higher temperature of $10^2\,T_0$. The computational box has a size of $(1.2\,\pc)^3$. We apply outflow boundary conditions for the magnetohydrodynamics and isolated boundaries for the self-gravity of the clump. We add a uniform magnetic field with $B_0=50\,\mu\Gauss$, initially pointing in the $z$-direction of our computational domain, consistent with the average magnetic field strength measured for clouds of this size and density \citep{HeilesTroland2005,CrutcherEtAl2010}. Finally, we add a turbulent velocity field.

\subsubsection{Turbulence} \label{sec:turb}

Turbulence is extremely important, because without it, our clump would collapse along the magnetic field lines to form a pancake with only a single star in the center, contrary to what is observed in real clouds. To make it more realistic and to seed initial perturbations from which dense cores can self-consistently form, we add a turbulent velocity field, similar to previous numerical studies of star cluster formation \citep[e.g.,][]{KlessenHeitschMacLow2000,HeitschMacLowKlessen2001,BateBonnellBromm2003,ClarkEtAl2005,KrumholzKleinMcKee2007,PriceBate2008,SmithClarkBonnell2008,FederrathBanerjeeClarkKlessen2010,GirichidisEtAl2011,MyersEtAl2013}. From sub-millimeter, molecular line observations, we know that clouds are turbulent and that the turbulent velocity dispersion on scale $\ell$ follows a power law,
\begin{equation} \label{eq:sigmav}
\sigma_v(\ell) = \sigma_V\,(\ell/L)^p \,,
\end{equation}
where $\sigma_V\sim 1\,\km\,\s^{-1}$ is the three-dimensional, non-thermal velocity dispersion on the scale $L\sim1\,\pc$, and $p\sim0.5$ from observations in Milky Way clouds \citep{Larson1981,SolomonEtAl1987,OssenkopfMacLow2002,HeyerBrunt2004,HeyerEtAl2009,RomanDuvalEtAl2011}. Note that this scaling of velocity with size is similar, but significantly different from the \citet{Kolmogorov1941c} scaling of incompressible turbulence, where $v\propto\ell^{1/3}$, i.e., $p\sim0.33$ instead of $0.5$. Indeed, high-resolution simulations of supersonic, compressible turbulence have confirmed the stronger scaling of the turbulent velocity with length scale in Equation~(\ref{eq:sigmav}) with $p=0.5$ \citep{KritsukEtAl2007,SchmidtEtAl2009,FederrathDuvalKlessenSchmidtMacLow2010,KonstandinEtAl2012,Federrath2013}, because molecular cloud turbulence is highly compressible and supersonic, unlike the incompressible Kolmogorov turbulence. According to the linewidth--size relation given by Equation~(\ref{eq:sigmav}) and our clump diameter $\ell=1\,\pc$, we set the turbulent velocity to $\sigma_v=1\,\km\,\s^{-1}$, corresponding to a turbulent Mach number $\mach=\sigma_v/\cs=5$, also consistent with the clump sample in \citet{NakamuraLi2014}.

Our initial turbulent velocity field is generated in Fourier space such that the power spectrum of velocity fluctuations follows the power law, $P \propto d\sigma_v^2/dk \propto k^{-2p-1} \propto k^{-2}$, thus obeying the observed velocity scaling given by Equation~(\ref{eq:sigmav}) with $p=0.5$. We do not drive the turbulence in this numerical experiment. The only remaining degree of freedom lies in the choice of how much power is distributed to the solenoidal (rotational or divergence-free) component of the velocity field, $\nabla\times\vect{v}$, relative to the power in the compressive (longitudinal or curl-free) component, $\nabla\cdot\vect{v}$. We can control the relative strength of the two components by performing a Helmholtz decomposition in Fourier space, resulting in the solenoidal component $P_\mathrm{sol}(k)$ and the compressive component $P_\mathrm{comp}(k)$. Note that the total power is always the sum of the two components, $P = P_\mathrm{sol}+P_\mathrm{comp}$. Following studies of driven turbulence employing this decomposition \citep{KritsukEtAl2007,FederrathKlessenSchmidt2008,SchmidtFederrathKlessen2008,SchmidtEtAl2009,Federrath2013}, we set the ratio
$\chi = P_\mathrm{sol} / (P_\mathrm{sol}+P_\mathrm{comp})$
to the natural mixture $\chi = 2/3$. The analogy of waves-like perturbation can be used to explain this natural ratio, because transverse waves occupy two of the three spatial dimensions and longitudinal waves occupy only one of the three, leading to the simple estimate $\chi = 2 / (2+1)$. This ratio enters the turbulence mode mixture, denoted as the $b$ parameter in the turbulent density variance--Mach number relation \citep{PadoanNordlund2011,MolinaEtAl2012,PadoanEtAl2014},
\begin{equation}
\sigs^2=\ln\left(1+b^2\mach^2\frac{\beta}{\beta+1}\right)\,.
\label{eq:sigs}
\end{equation}
A natural mixture of $\chi=2/3$ corresponds to $b=0.4$ \citep{FederrathDuvalKlessenSchmidtMacLow2010}. The turbulence mode mixture $b$ is important, because it has a profound effect on the star formation rate and efficiency \citep{GirichidisEtAl2011,FederrathKlessen2012,FederrathKlessen2013,PadoanEtAl2014}. In the numerical and theoretical studies by \citet{FederrathKlessenSchmidt2008,FederrathDuvalKlessenSchmidtMacLow2010}, $b$ was found to vary in the interval \mbox{$b=1/3\hdots1$}, with purely solenoidal driving of the turbulence producing $b=1/3$, purely compressive driving leading to $b=1$, and the natural mixture resulting in $b=0.4$, which we assume here. Observational studies also indicate that $b$ can take on a range of values, consistent with the theoretical interval. For instance, \citet{PadoanJonesNordlund1997} and \citet{Brunt2010} find $b\sim0.5$, as summarized in \citet{PriceFederrathBrunt2011}. \citet{KainulainenTan2013} and \citet{KainulainenFederrathHenning2013,KainulainenFederrathHenning2014} find \mbox{$b\sim0.3$--$0.5$} from a large sample of nearby molecular clouds. \citet{GinsburgFederrathDarling2013} measured a lower limit of $b>0.4$ for the GRSMC 43.30-0.33 cloud. The turbulence mixture parameter $b$ thus appears to vary in different clouds, which is likely a result of the different mechanisms that drive cloud turbulence \citep{MacLowKlessen2004,Elmegreen2009,FederrathKlessen2012,Krumholz2014}. Besides the measurements of $b$ based on the density probability distribution functions quoted above, a new observational method based on estimating the solenoidal-to-compressional ratio in the velocity field may also lead to a more direct measure of $b$ \citep{BruntFederrath2014}. For simplicity, we concentrate here on the average case, $b=0.4$, and leave a more detailed study of the influence of $b$ on the SFR and IMF for a future study.

\subsubsection{Dimensionless cloud parameters} \label{sec:cloudparams}

Given the molecular clump mass, radius, turbulent velocity dispersion, and magnetic field strength, the virial parameter, which is twice the ratio of kinetic to gravitational energy \citep{BertoldiMcKee1992,FederrathKlessen2012,PadoanEtAl2014} of the clump is
\begin{equation} \label{eq:alphavir}
\alphavir=2 E_\mathrm{kin} / E_\mathrm{grav} = 5 \sigma_v^2 R / (3G M_\mathrm{clump}) = 0.39\,.
\end{equation}
The mass-to-flux ratio is $5.5$, the magnetic to gravitational energy is $E_\mathrm{mag}/E_\mathrm{grav}=0.059$, and the ratio of thermal to magnetic pressure, called `plasma beta', is
\begin{equation} \label{eq:beta}
\beta = P_\mathrm{th} / P_\mathrm{mag} = 8\pi \rho_0\cs^2 / B_0^2 = 0.26.
\end{equation}
We will use these dimensionless energy ratios, as well as the turbulence Mach number $\mach=5$ and mode mixture parameter $b=0.4$ (see previous paragraph) in the next section, in order to calculate a theoretical prediction for the SFR of the clump for comparison with the SFR measured directly in the simulations.

\subsubsection{List of star cluster simulations}

Table~\ref{tab:clustersims} lists all star cluster formation simulations with and without outflow/jet feedback. We run the same simulations with a range of numerical resolutions. The highest-resolution star cluster calculation (run \texttt{Cluster\_4096\_SGSM} in Table~\ref{tab:clustersims}) has a maximum effective resolution of $4096^3$ grid cells, a minimum cell size of $60.4\,\AU$ and a sink particle density threshold of $1.19\times10^{-15}\,\g\,\cm^{-3}$, where the gas starts to become optically thick and slowly heats up (see Equations~\ref{eq:eos1} and~\ref{eq:eos2}), comparable to the lowest resolution of our SGS outflow test calculations above, which had a minimum cells size of $62.7\,\AU$ (see run \texttt{SGSM\_L07} in Table~\ref{tab:testsims}).

\subsection{Results}

\subsubsection{The star formation efficiency}

\begin{figure*}
\centerline{\includegraphics[width=1.0\linewidth]{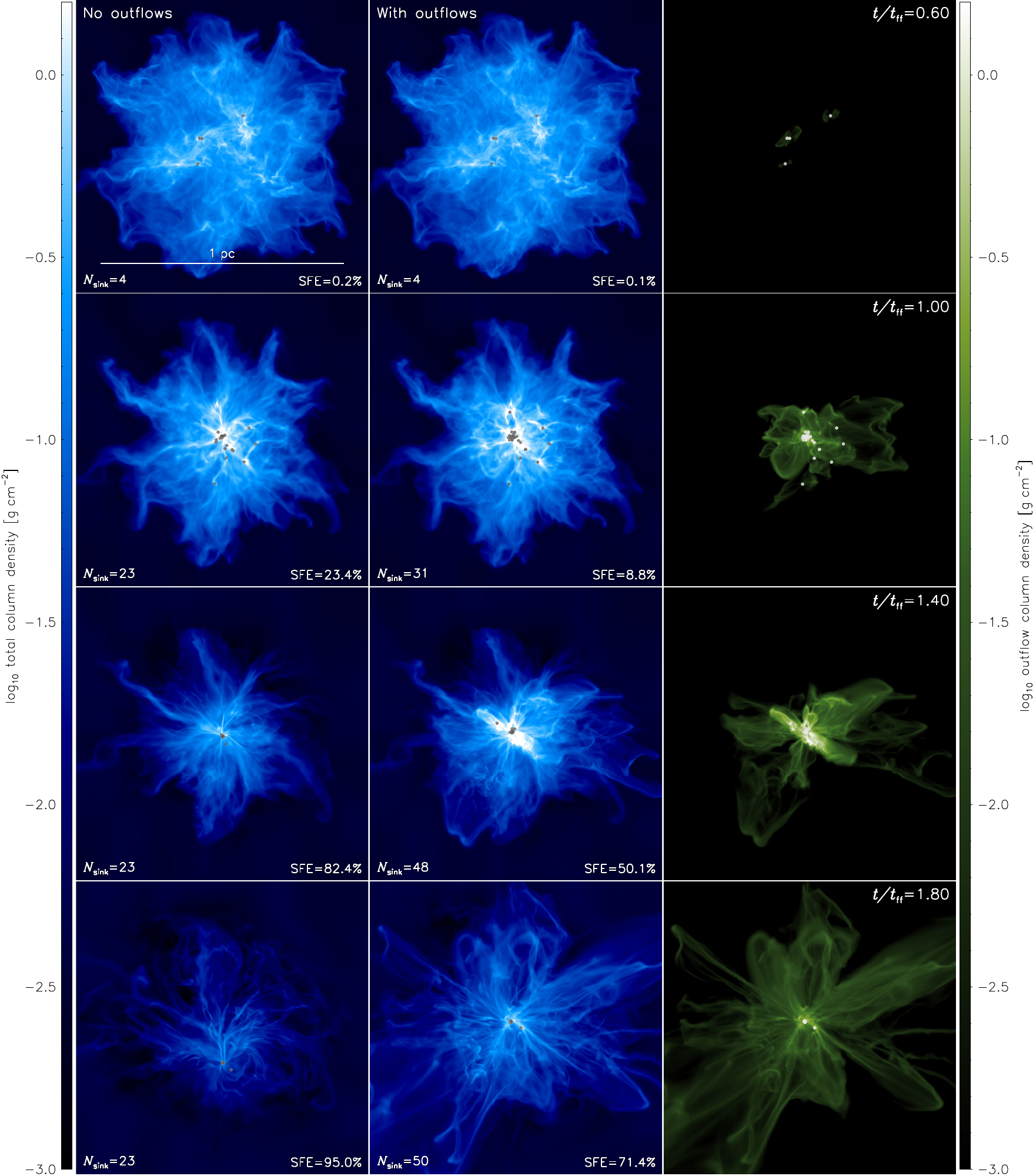}}
\caption{Time series of column density images for the control run without outflows (left-hand column) and with SGS outflow feedback activated (middle and right-hand columns) for times $t/\tff=0.6$, $1.0$, $1.4$, and $1.8$ (from top to bottom). The right-hand column shows the outflowing mass, i.e., the material that has been ejected by the sink particles through our SGS outflow model. At late times, the outflows and jets break out of the dense core and more than twice as many sink particles have formed compared to the control run without outflow feedback at $t=1.8\,\tff$. Thus, outflow feedback transforms the structure of the parent cloud and seems to have a strong impact on the SFR and on the characteristic mass of the IMF (to be quantified below). (\emph{Movies are available in the online version.})}
\label{fig:clusterimages}
\end{figure*}

Figure~\ref{fig:clusterimages} shows a time sequence of column density projections of the standard star cluster runs without outflows (\texttt{Cluster\_512\_NoSG}, left-hand panels) and with our SGS outflow model activated (\texttt{Cluster\_512\_SGSM}, middle panels). The right-hand panel additionally shows the column density of outflowing gas, which we marked using passive scalar advection, i.e., accreted gas that has been directly re-injected by our SGS outflow model. The top panels show the time $t=0.6\,\tff$, when the first four sink particles have just formed and started launching bipolar outflows. Two overlapping and nearly aligned outflows emerge from the first binary system with an initial separation of $\Delta (x, y, z) = (1.2, 0.9, 2.9)\times10^3\,\AU$, close to the center of the molecular clump. After one freefall time, $t=1.0\,\tff$ (2nd row), multiple stars and intersecting outflows have formed and some of the jets have already started to break out of the original molecular clump. At this time, 31 sinks have formed in the outflow run, while only 23 have formed in the run without outflows. The accreted mass is even more affected than the sink count, with $9\%$ of the original clump being accreted in the outflow run and more than three times as much ($23\%$) in the no-outflow run. In the following evolution ($t/\tff=1.4$ and 1.8), the outflows become stronger and break out of the boundary of our computational domain with high speed, leaving a very dense star cluster behind. While the entire clump is eventually accreted in the no-outflow run ($\sfe\to100\%$), the final $\sfe\lesssim75\%$ in the outflow feedback case\footnote{The star formation efficiency is defined as $\sfe=M_\star/(M_\star+M_\mathrm{gas})$ with the instantaneous star mass $M_\star$ and the instantaneous gas mass $M_\mathrm{gas}$ \citep[see, e.g., Equation~2 in][]{FederrathKlessen2013}.}. More than twice as many sink particles have formed compared to the no-outflow case at very late times. Already at $t=1.0\,\tff$ has the sink particle count increased by a factor of $\sim1.5$, if outflow feedback is included.

\begin{figure}
\centerline{\includegraphics[width=0.95\linewidth]{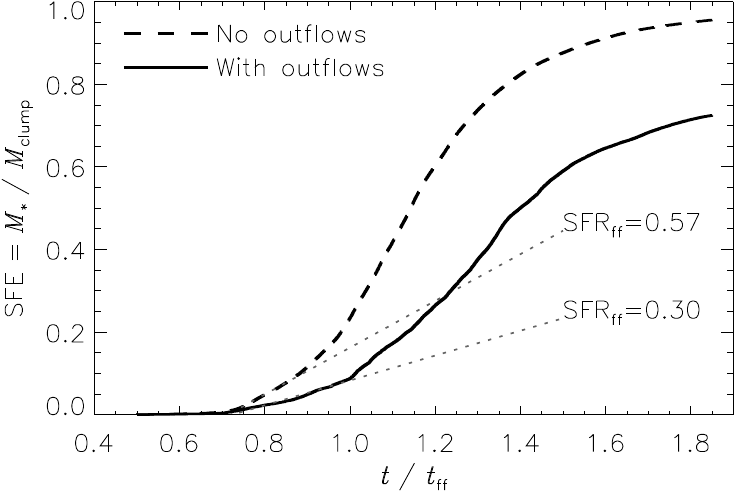}}
\caption{Time evolution of the star formation efficiency $\sfe=M_\star/M_\mathrm{clump}$ for the star cluster formation run without outflows (dashed line) and with outflow feedback (solid line). The star formation rate per freefall time ($\sfrff$) was determined by a linear fit in the interval \mbox{$\sfe=1\%$--$10\%$}. Outflow feedback reduces the $\sfrff$ by about a factor of two. Without outflows, all of the clump mass is eventually converted into stars, i.e., the SFE reaches 100\%. In contrast, the SFE only reaches about $75\%$ when outflow feedback in included. Thus, about $25\%$ of the original clump, i.e., more than $120\,\msol$, is ejected from the clump via high-speed jets and outflows (see also Figure~\ref{fig:clusterimages}).}
\label{fig:sfeevol}
\end{figure}

We now explore the time evolution and sink particle properties in more detail, with the ultimate goal of determining the effect of outflow feedback on the SFR and IMF. Figure~\ref{fig:sfeevol} shows the $\sfe$ as a function of time in units of the freefall time, Equation~(\ref{eq:tff}). We see that the $\sfe$ rises more slowly in the run with outflow feedback and that the curve begins to saturate toward $\sfe\sim75\%$. In contrast, when outflow feedback is ignored, the whole clump is eventually accreted onto stars. Thus, about $25\%$ of the original clump mass, i.e., more than $120\,\msol$, is ejected from the clump in high-speed jets and outflows, consistent with an observational analysis of the mass and momentum carried by outflows in Perseus \citep{ArceEtAl2010}. About $40\%$ of that mass ($\sim50\,\msol$) was accreted or re-accreted and then directly re-injected by our SGS outflow model, as measured with the passive scalar field shown in the right-hand panels of Figure~\ref{fig:clusterimages}. The mass fractions of injected and entrained material are thus similar to what we measured in the SGS test simulations in Section~\ref{sec:tests}.

We emphasize that our $\sfe\lesssim75\%$ with outflow feedback is an upper limit, i.e., the effect of outflows and jets would reduce the accreted mass even further in a more realistic setup that is not as gravitationally unstable as this one. Especially the virial parameter with $\alphavir\sim0.4$, is relatively low, the clump is isolated and cannot experience any shear or tidal forces from surrounding matter, and the turbulence is only driven from the inside. In contrast, real molecular clumps are embedded in a larger molecular cloud with turbulence driven from the outside by various different mechanisms \citep{MacLowKlessen2004,Elmegreen2009,KlessenHennebelle2010,FederrathKlessen2012}. We discuss these limitations further in Section~\ref{sec:limitations} below. Thus, the impact of outflows and jets in our setup is likely underestimated and the final $\sfe\lesssim75\%$ would be even lower in a more realistic setup. Indeed, dense cluster-forming regions have SFEs of up to $50\%$ \citep{WilkingLada1983,OlmiTesti2002} and eventually radiative and explosive feedback from massive stars and supernovae may leave a star cluster behind that is almost devoid of any gas ($\sfe\to1$). In contrast to radiative and explosive feedback from massive stars, however, outflow feedback is the first type of feedback to start and is also driven by low- and intermediate-mass stars \citep{KrumholzEtAl2014}. Thus, our results show that outflow and jet feedback may already lead to a significant reduction of the clump mass before other types of feedback can kick in.

\subsubsection{Star formation rate and core-to-star efficiency} \label{sec:sfr}

We now want to estimate the SFR. The slope of the curves in Figure~\ref{fig:sfeevol} provide a direct measure of the dimensionless SFR, called `star formation rate per freefall time' ($\sfrff$), because we plot the SFE versus time, in units of the freefall time, Equation~(\ref{eq:tff}). Using a linear fit within $\sfe=1\%$--$10\%$, we find slopes $\sfrff=0.57\pm0.16$ (no outflows) and $\sfrff=0.30\pm0.09$ (with outflows). The fit range was chosen to best represent the range of typical SFEs in a whole molecular cloud \citep{EvansEtAl2009,FederrathKlessen2013} and to allow us to directly compare with analytic SFR models below, by using the initial clump parameters (see Section~\ref{sec:cloudparams}) before they change significantly during the global collapse of the molecular clump. We note that the SFEs in dense, cluster-forming clumps can reach values of $\sim50\%$ \citep{WilkingLada1983,OlmiTesti2002}, reflecting the scale- and density-dependent hierarchy of SFEs \citep{FederrathKlessen2013}, eventually approaching the core-to-star efficiency. Such high SFEs are nearly compatible with the final $\sfe\sim75\%$ in our cluster-formation run with outflow feedback (see previous section), but still about $25\%$ too high, because our molecular clump has a very low virial parameter, is completely isolated from its parent molecular cloud, and we have not included other types of feedback, which might reduce the SFE further (see the discussion in Section~\ref{sec:limitations}).

The error bars of our $\sfrff$ measurements, $\sfrff=0.57\pm0.16$ (no outflows) and $\sfrff=0.30\pm0.09$ (with outflows), have been estimated by taking into account variations of the fit range within $\sfe=1\%$--$10\%$, systematic changes with numerical resolution, and statistical uncertainties introduced by the choice of random seed for the initial turbulence, with the latter being the dominant source of uncertainty. We see that outflow feedback reduces the SFR significantly and most likely by about a factor of two. But how does this compare to analytic predictions?

Theoretical models derive the $\sfrff$ from the statistics of supersonic, magnetized turbulence \citep{KrumholzMcKee2005,PadoanNordlund2011,HennebelleChabrier2011,FederrathKlessen2012}, by integrating the density probability distribution function (PDF) \citep[e.g.,][]{Vazquez1994,FederrathKlessenSchmidt2008,MoraghanKimYoon2013} from a threshold density to infinity. The PDF is sensitive to the strength and driving of the turbulence \citep{FederrathKlessenSchmidt2008,PriceFederrathBrunt2011} and the density threshold is derived by comparing the sonic scale \citep{VazquezBallesterosKlessen2003,FederrathDuvalKlessenSchmidtMacLow2010} with the Jeans scale \citep{PadoanEtAl2014}. Taken all together, this leads to a theoretical prediction for the SFR \citep[see the derivations in][]{FederrathKlessen2012,HennebelleChabrier2013,Federrath2013sflaw}:

\begin{equation} \label{eq:sfrff}
\sfrff=\frac{\eps}{2\phit}\exp\left(\frac{3}{8}\sigs^2\right)\left[1+\mathrm{erf}\left(\frac{\sigs^2-s_\mathrm{crit}}{\sqrt{2\sigs^2}}\right)\right].
\end{equation}

The parameter $\eps$ is the core-to-star efficiency, which we want to estimate below, and $\phit$ is a fudge factor of order unity. The standard deviation $\sigs$ of the PDF is given by Equation~(\ref{eq:sigs}) and depends on the Mach number $\mach$, on the turbulence mode-mixture parameter $b$, and on plasma $\beta$. The density threshold $\scrit$ in Equation~(\ref{eq:sfrff}) is derived using different assumptions in \citet[][]{KrumholzMcKee2005} (hereafter `KM'), \citet[][]{PadoanNordlund2011} (hereafter `PN'), and \citet[][]{HennebelleChabrier2011} (hereafter `HC') and depends additionally on the virial parameter, $\alphavir$ (Equation~\ref{eq:alphavir}). Thus, the theoretical SFR depends on four basic cloud parameters: $\alphavir$, $\mach$, $b$, and $\beta$. In total, there are six different flavors of the turbulence-regulated theory of the SFR: the original KM, PN, and HC theories and the multi-freefall versions defined in \citet{HennebelleChabrier2011} and extended and summarized in \citet{FederrathKlessen2012}.

Given the four input parameters of our initial molecular clump, $\alphavir=0.39$ (Equation~\ref{eq:alphavir}), $\mach=5.0$, $b=0.4$ (see Section~\ref{sec:turb}), and $\beta=0.26$ (Equation~\ref{eq:beta}), we can compute predictions of the $\sfrff$ from each of the six theories. We note that some of the original theories (KM and HC) did not include magnetic fields, which were later added by \citet{FederrathKlessen2012}. We thus also consider all theoretical cases for $\beta\to\infty$ (no magnetic fields), in order to see the effect of adding magnetic pressure.

\begin{table*}
\caption{Theoretical predictions for the star formation rate per freefall time ($\sfrff$) in our isolated star cluster.}
\label{tab:theories}
\def\arraystretch{1.3}
\begin{tabular*}{\linewidth}{@{\extracolsep{\fill} }lcccccc}
\hline
\hline
 & \multicolumn{3}{c}{\mbox{Original Theories}} & \multicolumn{3}{c}{\mbox{Multi-freefall Theories}} \\
 & KM & PN & HC & Multi-ff KM & Multi-ff PN & Multi-ff HC \\ 
\hline
$\sfrff$ for $\beta\to\infty$ (HD) & $0.016$ & $0.28$ & $0.084$ & $0.81$ & $0.81$ & $0.37$ \\
$\sfrff$ for $\beta=0.26$ (MHD) & $0.024$ & $0.33$ & $0.00023$ & $0.58$ & $0.59$ & $0.25$ \\
\hline
\end{tabular*}
\\
\textbf{Notes.} The original theories by KM, PN, and HC were evaluated with their respective favorite parameter choices: $\eps/\phit(\mathrm{KM,PN,HC})$=$(0.52,0.5,0.1$) and $\fudge(\mathrm{KM,PN,HC})$=$(1.12, 0.35, 0.1)$. For the multi-freefall versions, we use the best-fit theory parameters determined in \citet{FederrathKlessen2012}: $\eps/\phit(\mathrm{KM,PN,HC})$=$(0.46,0.47,0.2$) and $\fudge(\mathrm{KM,PN,HC})$=$(0.17, 1.0, 5.9)$. The SFR in the simulations is $\sfrff=0.57\pm0.16$ without outflows and $\sfrff=0.30\pm0.09$ with outflow feedback. The multi-freefall theories were all calibrated with non-feedback simulations and a comparison with observations yielded a feedback correction factor with the core-to-star efficiency $\eps\sim0.5$ \citep{FederrathKlessen2012}, consistent with the factor $\sim2$ lower $\sfrff$ when outflow feedback is included.
\end{table*}

Table~\ref{tab:theories} lists all the theoretical predictions of the SFR for our star cluster. Note that the standard parameters of the original theories \citep[$\phit$, $\phix$ $\theta$, and $\ycut$; see][]{FederrathKlessen2012} were taken from the respective original theory (KM, PN, HC) and the parameters for the multi-ff KM, multi-ff PN, and multi-ff HC theories were taken from the best-fit theory--simulation comparison using a comprehensive set of cloud simulations in \citet{FederrathKlessen2012}. They all assume that the core-to-star efficiency $\eps=1$, because the theories cannot predict this value from first principles and the simulations used to determine the theory parameters did not include any type of feedback. Thus, the best direct comparison of the theoretical predictions is with our no-outflow run, including magnetic fields. Consistent with the conclusions in \citet{FederrathKlessen2012}, we find that the multi-ff KM and the multi-ff PN models provide the best theoretical predictions, with $\sfrff=0.58$ and $\sfrff=0.59$, respectively, both in agreement with the simulation measurement ($\sfrff=0.57\pm0.16$). The multi-ff HC and the original PN theory underestimate the true SFR by a factor of two. The other original theories tend to underestimate the simulation SFR even more. Comparing the purely hydrodynamic (HD) theory case ($\beta\to\infty$) with the MHD theory case (using our simulation $\beta=0.26$), we find that all multi-freefall theories and the original HC theory \citep[with magnetic pressure effects added by][]{FederrathKlessen2012} correctly predict a reduction of the SFR when magnetic pressure is accounted for, while the original KM and PN theories predict an increased SFR when magnetic fields are included. We conclude that the multi-freefall KM and the multi-freefall PN models with the parameters derived in \citet{FederrathKlessen2012} currently provide the best theoretical predictions for the SFR.

Finally, we see in Figure~\ref{fig:sfeevol} that outflow feedback reduces the SFR by a factor of about two in our simulations. This directly relates to the core-to-star efficiency $\eps$ and we conclude that $\eps\sim0.5$ when outflow feedback is included. This value of $\eps$ is in agreement with theoretical work by \citet{MatznerMcKee2000}, with isolated single-star simulations \citep{MachidaHosokawa2013}, and with the comparison of numerical simulations with Milky Way observations by \citet{HeidermanEtAl2010} in \citet{FederrathKlessen2012}. However, $\eps$ may be even smaller if other feedback effects were accounted for, e.g., radiation feedback, which we have not included. On the other hand, radiation feedback from low- and intermediate-mass stars is primarily a type of thermal feedback. It leads to an increased gas temperature by a factor of 2--3 locally within $\lesssim1000\,\AU$ around each protostar \citep{Bate2009rad,OffnerEtAl2009}, potentially reducing or even suppressing fragmentation on smaller scales. However, radiation from low- and intermediate-mass stars hardly affects the momentum balance and dynamics. We also note that both \citet{Bate2009rad} and \citet{Bate2012} show that thermal feedback has little effect on the overall star formation rate and efficiency of a star cluster. In contrast, jets and outflows are the most important type of mechanical feedback in this context \citep[][]{KrumholzEtAl2014}, capable of ejecting material from the star-forming molecular clump and affecting regions that are more than a parsec away from the protostar.

\subsubsection{The impact of outflow and jet feedback on the characteristic mass of the IMF} \label{sec:IMF}

\begin{figure}
\centerline{\includegraphics[width=0.95\linewidth]{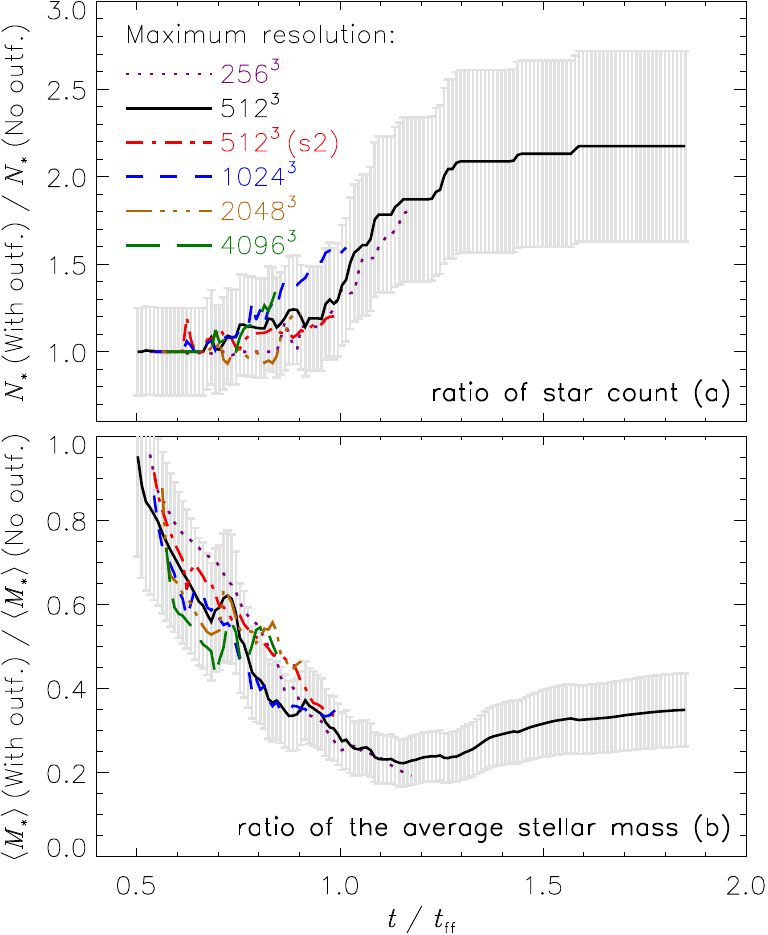}}
\caption{Time evolution of the star count ratio $N_\star(\mathrm{With\;outflows}) / N_\star(\mathrm{No\;outflows})$ (panel a) and ratio of the average stellar mass $\langle M_\star \rangle (\mathrm{With\;outflows}) / \langle M_\star \rangle (\mathrm{No\;outflows})$ (panel b) for different maximum, effective resolutions of $256^3$--$4096^3$ (see Table~\ref{tab:clustersims}). The dash-dotted lines show the same as our standard cluster run, but initialized with a different random seed for the turbulence (labelled \mbox{`$512^3\;(\mathrm{s}2)$'}). After one freefall time, outflow feedback has increased the star count by a factor of $\sim1.5$ and reduced the average mass of stars by a factor of $\sim3$. The gray error bars indicate a 25\% uncertainty level (shown only for the standard runs with $512^3$ resolution).}
\label{fig:sinksevol}
\end{figure}

Considering the significant reduction of the SFR and core-to-star efficiency caused by outflow feedback, we expect a strong impact on the stellar mass and thus on the characteristic mass of the IMF. To quantify this, we plot in the top panel of Figure~\ref{fig:sinksevol} the number of sink particles formed in the run with outflow feedback divided by the number of sinks formed in the no-outflow run, as a function of time. We see that $N_\star(\mathrm{With\;outflows})/N_\star(\mathrm{No\;outflows}) \sim 1.5$ after one freefall time. The reason for the increased star count is that the jets and outflows break the existing filamentary accretion flows into multiple streams. They perturb the local density structure such that multiple new cores and stars form \citep[see also][]{WangEtAl2010}. We note that this mechanism leading to the increased star count due to outflow feedback is hardly affected by other forms of feedback, such as radiation feedback \citep{HansenEtAl2012}.

The sink count ratio hardly depends on the numerical resolution, as we can see from the resolution study with maximum, effective resolutions of $256^3$--$4096^3$ in Figure~\ref{fig:sinksevol}. These resolutions correspond to the smallest AMR cell sizes of $\dx=970\,\AU$ down to $60\,\AU$ (see Table~\ref{tab:clustersims}). Although the absolute level of fragmentation depends on the numerical resolution (see Section~\ref{sec:limitations}), the star count ratio is almost insensitive to the resolution and shows that after about one freefall time, about $1.5$ times as many stars have formed, when outflow feedback is included. This result is also robust against changes in the random seed used to initialize the turbulent velocity field (see Section~\ref{sec:turb}), as shown by an additional model labelled \mbox{`$512^3\;(\mathrm{s}2)$'} in Figure~\ref{fig:sinksevol}. Thus, we find that outflows lead to a significantly increased star count, consistent with previous numerical experiments by \citet{LiEtAl2010} and \citet{HansenEtAl2012}, who also observed an increased star count by a factor of $\sim1.5$ when outflow feedback was included.

Finally, we determine the impact of outflows and jets on the average star mass. In analogy to the star count ratio, the bottom panel of Figure~\ref{fig:sinksevol} shows the ratio of the average stellar mass, $\langle M_\star\rangle(\mathrm{With\;outf.})/\langle M_\star\rangle(\mathrm{No\;outf.})$. This ratio gradually decreases, almost independent of the numerical resolution and turbulent seed, and saturates at about 1/3 after one freefall time. Previous numerical studies also found ratios of 1/2--1/3 \citep{LiEtAl2010,HansenEtAl2012}. Also \citet{Krumholz2011} assumes a factor of $(1/2)^{13/9}\sim0.37$ caused by outflow feedback, when he derives the characteristic mass of stars from fundamental physical constants. Here we measure this ratio directly in MHD simulations and show that outflow feedback does indeed reduce the characteristic mass of stars by a factor of $\sim3$ (see Figure~\ref{fig:sinksevol}), suggesting that it might help to explain the observed systematic shift of the core mass function to the IMF by about the same factor, $0.3$--$0.4$ \citep{AlvesLombardiLada2007,NutterWardThompson2007,EnochEtAl2008,Myers2008,AndreEtAl2010,KoenyvesEtAl2010,OffnerEtAl2014,FrankEtAl2014}. This could be interpreted and parametrized as an effective core-to-star efficiency $\eps_\mathrm{effective}\sim1/3$. Most importantly, we find here that it is the \emph{combined effect} of \emph{increased star count} by about a factor of $\sim1.5$ and \emph{decreased actual core-to-star efficiency} \mbox{$\eps\sim0.5$} (i.e., the fraction of material being ejected from the protostellar envelope) that leads to the stronger reduction of the characteristic star mass by
\begin{equation} \label{eq:eps_effective}
\eps_\mathrm{effective} = \eps \; \frac{N_\star(\mathrm{No\;outflows})}{N_\star(\mathrm{With\;outflows})} \sim \frac{1}{2}\times\frac{2}{3}=\frac{1}{3},
\end{equation}
than any of the two effects alone. Note the difference between the one-core to one-star efficiency denoted $\eps$ (used in e.g., Equation~\ref{eq:sfrff}), and the definition of the \emph{effective} core-to-star efficiency in Equation~(\ref{eq:eps_effective}), which describes the combined effect of $\eps\sim0.5$ and increased star count by $\sim1.5$. Thus, outflow feedback triggers additional fragmentation that reduces the average star mass further than $\eps$ alone. This result demonstrates that outflow feedback may be an essential ingredient for understanding the characteristic mass of the IMF.

\subsection{Caveats and limitations} \label{sec:limitations}

\subsubsection{Limited resolution and missing physics}

While our results of the star count ratio and average star mass ratio in Figure~\ref{fig:sinksevol} are converged with resolution, we caution that the \emph{absolute} star count and average star mass are not fully converged. This has two reasons: (1) our resolution is insufficient to fully resolve the opacity limit, and (2) we use a piecewise polytropic equation of state (EOS), Equations~(\ref{eq:eos1}) and~(\ref{eq:eos2}), instead of full radiation-hydrodynamics (cf.~Section~\ref{sec:methods}). While the first caveat leads us to underestimate the star count, the second tends to lead to an overestimate. Detailed radiation-hydrodynamical calculations have shown that the star count and the average star mass depend on the heating and cooling balance \citep{KrumholzKleinMcKee2007,Bate2009rad,CommerconEtAl2010}, although \citet{HansenEtAl2012} find that outflow feedback is crucial in determining the effect of radiation feedback. This is because outflow feedback comes first, drives cavities into the dense gas and reduces the average stellar mass, all leading to a significantly reduced effect of radiation when outflow feedback is included. Our primary goal here is to determine the role of outflow feedback for the SFR, SFE and IMF, and indeed, the relative star count and the relative average mass between our simulations with and without outflow feedback are robust, as demonstrated in Figure~\ref{fig:sinksevol}. The SFR and SFE are also converged already at moderate resolutions, as shown in dedicated resolution studies by \citet{PadoanNordlund2011} and \citet{FederrathKlessen2012}.

\subsubsection{Boundary and initial conditions} 

The initial and boundary conditions of the star-forming molecular clump are only rough approximations to real clouds. We start with an initial turbulent velocity field that is free to decay and is superimposed on a uniform density sphere. It is a conventional type of setup often used in numerical studies of star formation \citep[e.g.,][]{BateBonnellBromm2003,ClarkEtAl2005,KrumholzKleinMcKee2007,PriceBate2008,PriceBate2009,SmithClarkBonnell2008,FederrathBanerjeeClarkKlessen2010,WalchEtAl2010,GirichidisEtAl2011,MyersEtAl2014}. But it is unrealistic and leaves open the question of the origin of the turbulent velocity field. Given the importance of the turbulence for star formation \citep[][Section~\ref{sec:turb}]{MacLowKlessen2004,ElmegreenScalo2004,McKeeOstriker2007,HennebelleFalgarone2012,FederrathKlessen2012,Krumholz2014,KrumholzEtAl2014,PadoanEtAl2014}, this is a serious limitation of this type of setup.

In the simulations, the molecular clump is assumed to be completely disconnect from the larger molecular cloud in which it actually resides. But such a dense clump likely forms from large-scale turbulent compression modes inside a larger cloud \citep[e.g.,][]{HeyerBrunt2004,RomanDuvalEtAl2011}. These large-scale modes break up into smaller turbulent perturbations and continuously feed turbulent energy to the clump. Thus, the turbulence is more likely driven than decaying \citep[e.g.,][]{BruntHeyerMacLow2009,KlessenHennebelle2010}. Moreover, tidal forces from the matter surrounding the clump may effectively decrease the gravitational binding energy of the clump and increase the effective virial parameter \citep[][]{FederrathKlessen2012}. Real clumps are likely to be less gravitationally unstable than in our numerical setup and the effect of outflows and jets may be even larger than we estimate here, because they would be able to break out of the clump much more easily and transfer mass, momentum and angular momentum to other parts of the parent cloud.

\section{Summary and conclusions} \label{sec:conclusions}

We implemented and tested a new subgrid-scale (SGS) model for launching outflows and jets in MHD simulations of star cluster formation. Such an SGS outflow model is required, because the computational power currently available is orders of magnitude too small to solve this problem \emph{ab initio}. Our SGS model is the first to include angular momentum transfer, needed to solve the angular momentum problem. It is also the first to demonstrate convergence and reproduces the mass, momentum and angular momentum injection, as well as the speeds of real jets and outflows (cf.~Figures~\ref{fig:lrefimages} and~\ref{fig:totals_lref}). Our SGS outflow model is adaptive in the sense that the outflow properties on large scales do not depend significantly on the numerical resolution, when the SGS model is activated (cf.~Figure~\ref{fig:totals_sgs}). Even the best previous SGS outflow models likely underestimated the jet speeds, either because the jet component was not included in the model or the SGS launching region was not sufficiently resolved, as shown in our test calculations (cf.~Figure~\ref{fig:totals_rout}). Our model produces self-regulated feedback, shown by the fact that the outflow properties do not depend significantly on whether the mass transfer fraction is chosen to be $10\%$ or $30\%$ (cf.~Figure~\ref{fig:totals_misc}).

In Section~\ref{sec:cluster}, our new SGS outflow model is applied to the collapse of a magnetized, turbulent molecular clump, forming a star cluster. By comparing to an otherwise equivalent run without outflows, we draw the following conclusions:

\begin{itemize}

\item Multiple jets break out of the molecular clump, ejecting at least one quarter of the clump mass at high speed (cf.~Figure~\ref{fig:clusterimages}). These jets quickly reach distances greater than a parsec and may have a significant influence on the structure, dynamics and star formation in other parts of the parent molecular cloud.

\item Outflow feedback reduces the SFR by about a factor of two. Theoretical models based on the statistics of MHD turbulence correctly predict the SFR in the initial phase of star formation, before the clump starts to collapse globally (cf.~Figure~\ref{fig:sfeevol}).

\item The number of stars increases by a factor of $\sim1.5$ when outflow feedback is included and the average mass of stars decreases by a factor of $\sim3$. This reduced star mass is the result of the \emph{combination} of reduced accretion onto each star by a factor of $\eps\sim0.5$ and the increased star count, which leads to an effective core-to-star efficiency, $\eps_\mathrm{effective}\sim0.5/1.5\sim1/3$  (cf.~Equation~\ref{eq:eps_effective} and~Figure~\ref{fig:sinksevol}). Thus, jet and outflow feedback might be a crucial ingredient for understanding the characteristic mass of the IMF.

\end{itemize}

We conclude that outflow feedback has a dual role. On one hand, it limits accretion directly by removing mass from the core and indirectly by driving turbulence and entraining material that would have been accreted before it can reach the star-forming core. On the other hand, jets and outflows may also trigger star formation by driving local turbulent compressions in other parts of the same cloud and by tearing filaments apart, which disrupts the original, coherent accretion flow onto a single star and redirects it into multiple accretion channels that lead to the formation of multiple stars instead of a single one. Outflow feedback thus leads to a higher star count and to a significantly lower average star mass.

We additionally show in Appendix~\ref{app:jeans} that a Jeans resolution of at least 32 grid cells is required for convergence, so we propose $\jres\geq32\,\dx$ as the standard Jeans resolution criterion for hydrodynamic and MHD simulations involving self-gravitational collapse.

\acknowledgements
We thank M.~Bate, B.~Commercon, A.~Cunningham, P.~Girichidis, R.~Klein, P.~Kroupa, M.~Krumholz, C.~Matzner, C.~McKee, S.~Offner, R.~Parkin, T.~Peters, D.~Price, R.~Pudritz, and D.~Seifried for stimulating discussions about jets and outflows, and for comments on the paper.
CF acknowledges funding provided by the Australian Research Council's Discovery Projects (grant no.~DP110102191).
MS acknowledges kind support from Dieter Breitschwerdt (Dept.~of Astronomy \& Astrophysics, Technical University Berlin).
RB acknowledges support from the DFG through the grants BA 3706/1-1, BA 3706/3-1, BA 3706/4-1 as well as support through the SFB 676.
RSK acknowledges support from the Deutsche Forschungsgemeinschaft (DFG) via the SFB 881 (subproject B1, B2 and B5) ``The Milky Way System'', and the SPP (priority program) 1573 ``Physics of the ISM''. RSK also acknowledges support from the European Research Council under the European Community's Seventh Framework Programme (FP7/2007-2013) via the ERC Advanced Grant STARLIGHT (project number 339177).
We thank for supercomputing time provided by the Leibniz Rechenzentrum and the Gauss Centre for Supercomputing (grant pr32lo and PRACE grant pr89mu) and by the J\"ulich Supercomputing Centre (grant hhd20).
The software used in this work was in part developed by the DOE-supported Flash Center for Computational Science at the University of Chicago. This research was supported in part by the National Science Foundation under Grant No.~NSF PHY11-25915 during CF's visit to the Kavli Institute for Theoretical Physics, where this work was completed.

\appendix

\section{Jeans resolution study} \label{app:jeans}

All state-of-the-art numerical studies of star formation resolve the Jeans length during local collapse by more than four grid cells, $\jres\geq4\,\dx$, to avoid artificial fragmentation \citep{TrueloveEtAl1997}. The corresponding resolution criterion in smoothed particle hydrodynamics (SPH) simulations is discussed in \citet{BateBurkert1997}. In practice, most state-of-the-art simulations run with $4\leq\jres/\dx\lesssim8$ \citep[e.g.,][]{KrumholzKleinMcKee2012,HansenEtAl2012,SeifriedEtAl2012,MyersEtAl2013,MachidaHosokawa2013,OffnerArce2014}, because using more cells per Jeans length severely increases the computational cost. Although using $4\leq\jres/\dx\lesssim8$ safely prevents artificial fragmentation, the question is whether it is sufficient to resolve quantities such as the energy, momentum, angular momentum or magnetic field amplification during collapse.

\begin{figure*}
\centerline{\includegraphics[width=0.85\linewidth]{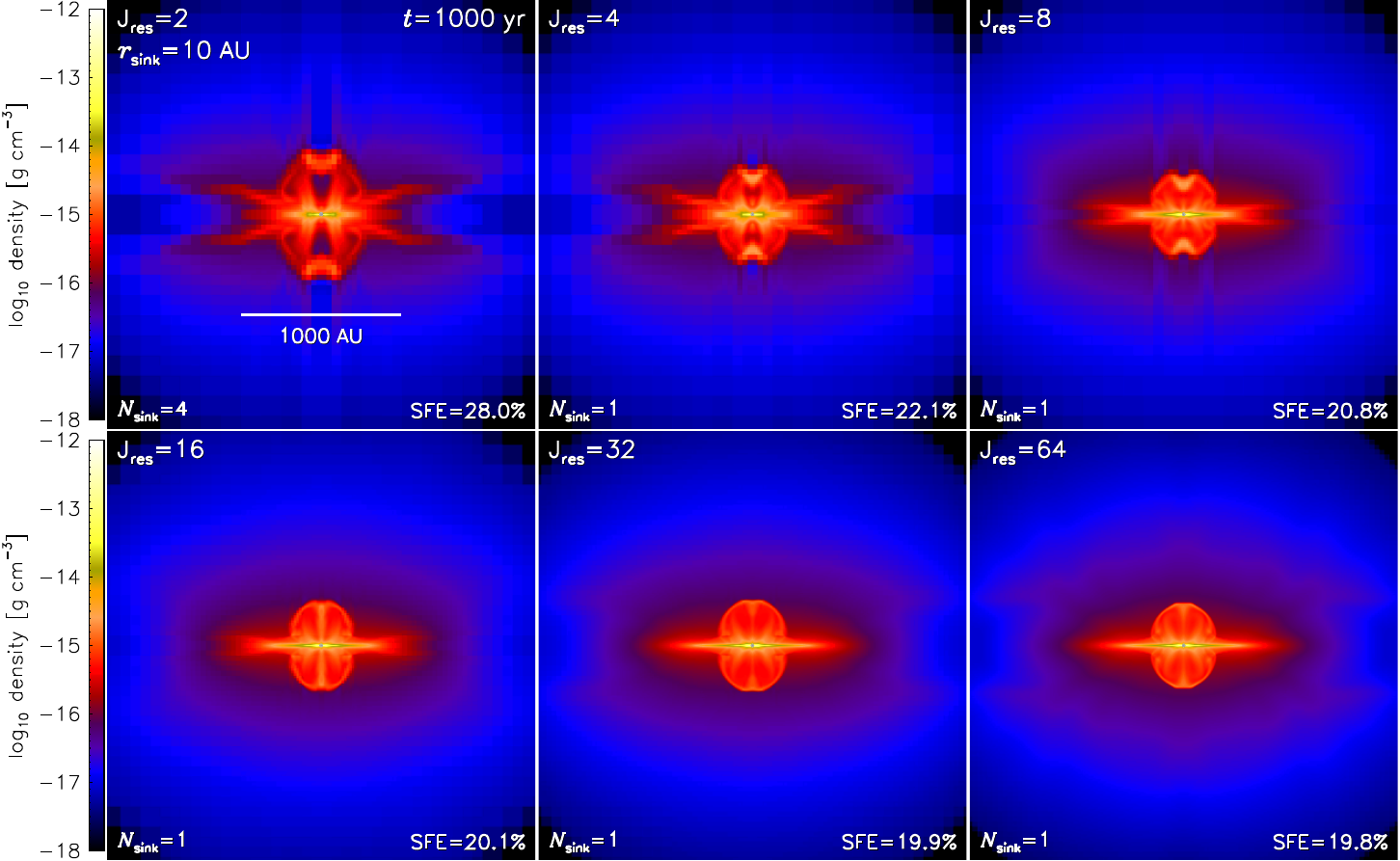}}
\caption{Jeans resolution study of disk and outflow formation. The panels show simulations \#16--19, \#2, and \#20 from Table~\ref{tab:testsims} with Jeans resolution $\jres/\dx=2$, $4$, $8$, $16$, $32$, and $64$, respectively. For $\jres/\dx=2$, we find artificial fragmentation into four sink particles, while only a single star forms for $\jres/\dx\geq4$ \citep[see also][]{TrueloveEtAl1997}. The disk and outflow structure, however, as well as the accretion rate (indicated by the $\sfe$ in the bottom-right corner of each frame) are only converged for $\jres/\dx\gtrsim32$, which is why we use this as the standard Jeans resolution criterion in all our production runs \citep[see also][]{FederrathSurSchleicherBanerjeeKlessen2011}.}
\label{fig:images_jres}
\end{figure*}

\begin{figure*}
\centerline{\includegraphics[width=0.92\linewidth]{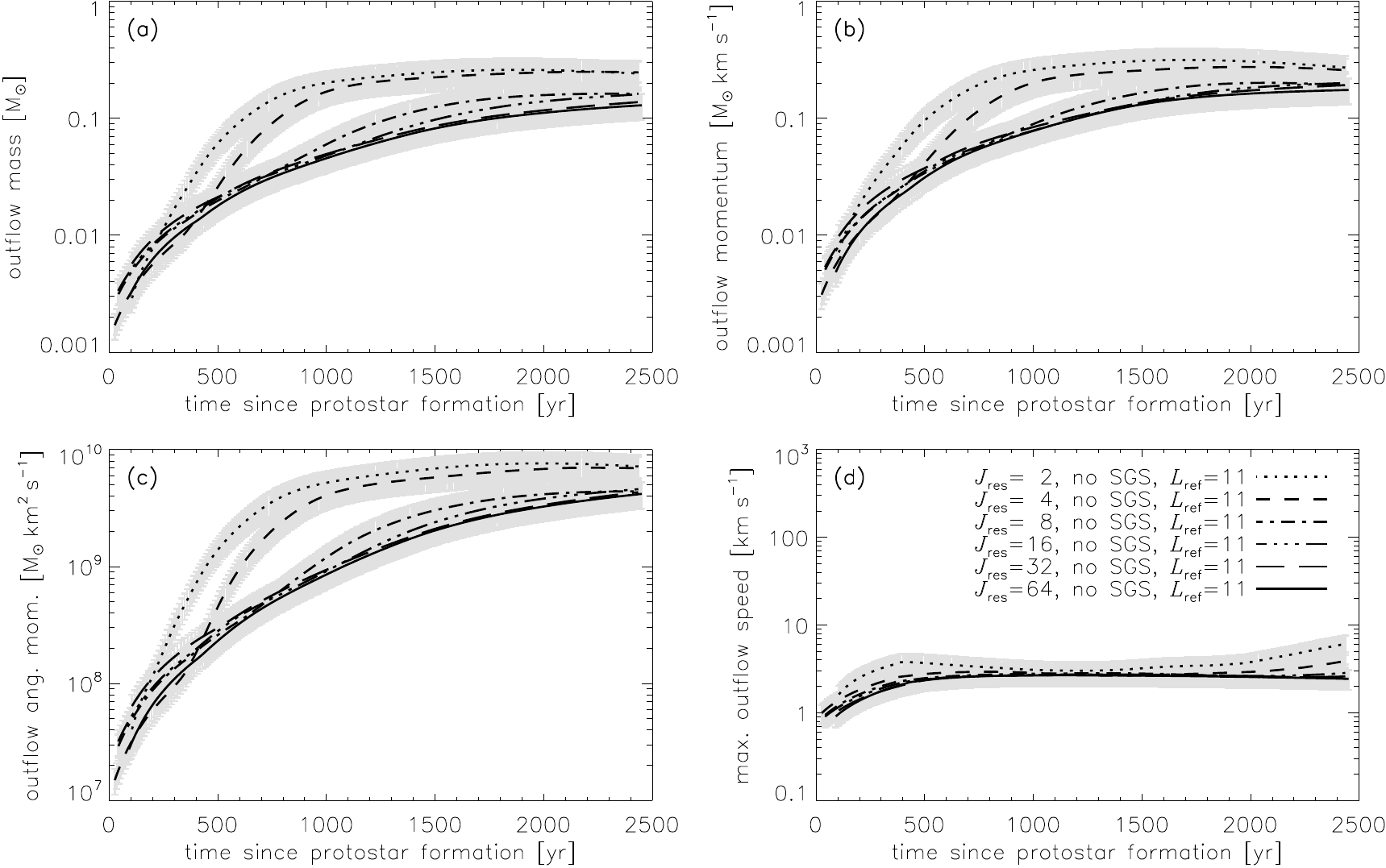}}
\caption{Same as Figure~\ref{fig:totals_lref}, but for calculations without SGS model, but different Jeans resolution $\jres/\dx=2$, $4$, $8$, $16$, $32$, and $64$. For convergence of all relevant quantities including the angular momentum, $\jres/\dx\gtrsim32$ is required \citep[see also][]{FederrathSurSchleicherBanerjeeKlessen2011}. Here, all outflow properties were determined by selecting all cells with an outflowing $z$-velocity component, located one scale height ($H=25\,\AU$) above and below the disk.}
\label{fig:totals_jres}
\end{figure*}

In order to quantify the effect of the Jeans resolution, we run test simulations \#16--19, \#2, and \#20 from Table~\ref{tab:testsims} (no SGS outflow model) with Jeans resolution $\jres/\dx=2$, $4$, $8$, $16$, $32$, and $64$, respectively. Density slices through the disk are shown in Figure~\ref{fig:images_jres}. Note that the outflow in these tests is not resolved, but we can still compare the effect of varying the Jeans resolution by comparing the otherwise identical runs with one another. We confirm the result in \citet{TrueloveEtAl1997} that $\jres=2\,\dx$ leads to artificial fragmentation, here into four sink particles, while for $\jres\geq4\,\dx$, only a single sink particle forms. However, the disk exhibits artificial flaring for $\jres\leq16\,\dx$. The $\sfe$ is overestimated by $11\%$ and $5\%$, respectively, for $\jres=4\,\dx$ and $8\,\dx$, while it is converged to within 0.5\% for $\jres\ge32\,\dx$.

Figure~\ref{fig:totals_jres} shows the outflow mass, momentum, angular momentum and speed as in Figure~\ref{fig:totals_lref}, but here for the different Jeans resolutions. Since this Jeans resolution study was run without the SGS model, the outflows are quite slow and do not reach sufficient distance from the disk to analyze them at $500\,\AU$ above and below the disk. We therefore select here all cells with a vertical position of one scale height $H\sim\cs\Omega^{-1}\sim\cs v_\phi^{-1}R = 25\,\AU$ above and below the disk midplane, $\left|z\right| > H$ and with an outflowing vertical velocity component, as before. We see that all outflow properties are sensitive to the Jeans resolution, if $\jres\leq16\,\dx$. Runs with $\jres=2\,\dx$ and $4\,\dx$ show strong qualitative differences, but runs with $8\leq\jres/\dx\leq16$ are already quite close to the runs with $\jres\geq32\,\dx$, in particular at late times. All relevant quantities are fully converged with $\jres\geq32\,\dx$. This is in agreement with the independent Jeans resolution studies by \citet{SurEtAl2010}, \citet{FederrathSurSchleicherBanerjeeKlessen2011}, and \citet{TurkEtAl2012} in the context of early-universe star formation. Despite the different context, dimension and physics, the numerical resolution criterion of $\jres\ge32\,\dx$ is virtually identical. We thus use the Jeans resolution criterion $\jres\geq32\,\dx$ throughout, for the SGS test simulations in Section~\ref{sec:tests} and for the star cluster production runs in Section~\ref{sec:cluster}. We suggest that all state-of-the-art hydrodynamic and MHD simulations involving gravitational collapse should employ this Jeans resolution criterion as the standard criterion.

\end{document}